\documentclass[a4paper,11pt]{article}
\pdfoutput=1 

\usepackage{jheppub} 
\usepackage[mathscr]{euscript}
\usepackage[T1]{fontenc} 

 \usepackage{musicography}
\usepackage{empheq}
  \usepackage{mathrsfs}
 
\usepackage{bm}
\usepackage{verbatim}
 
 \usepackage{makecell}
 \usepackage{float}
 \usepackage{hyperref}

 \usepackage[makeroom]{cancel}
 
 \newcommand{\bv}{ \begin{verbatim}}

    \newcommand{\bz}{{ \bar z}}
  
    \newcommand{\Split}{{ \mathsf{Split}}}

\newcommand{\be}{\begin{equation}}
\newcommand{\ee}{\end{equation}}
\newcommand{\bpm}{\begin{pmatrix}}
\newcommand{\epm}{\end{pmatrix}}

\newcommand{\PBK}[1]{\ensuremath{\begin{pmatrix}#1\end{pmatrix}}}

\newcommand{\EV}[1]{\langle #1 \rangle}
\newcommand{\beqn}{\begin{eqnarray}}
\newcommand{\eeqn}{\end{eqnarray}}

\usepackage{slashed}

\newcommand{\cT}{\mathcal T}
\newcommand{\cJ}{\mathcal J}
\newcommand{\cO}{\mathcal O}
\newcommand{\cR}{\mathcal R}

\newcommand{\cQ}{\mathcal Q}

\newcommand{\cH}{\mathcal H}
\newcommand{\cK}{\mathcal K}
\newcommand{\cL}{\mathcal L}
\newcommand{\cI}{\mathcal I}
\newcommand{\cN}{\mathcal N}

\newcommand{\sO}{\mathscr {O}}

\newcommand{\p}{\partial}

\DeclareMathOperator{\Real}{Re}

\author[\natural]{Hongliang Jiang }

 \affiliation[\natural{}]{Centre for Research in String Theory, School of Physics and Astronomy,
Queen Mary University of London, Mile End Road, E1 4NS,  UK}

 \emailAdd{h.jiang@qmul.ac.uk}

\preprint{QMUL-PH-21-36}

\title{\boldmath\huge Holographic Chiral Algebra: Supersymmetry, Infinite Ward Identities, and EFTs}

\abstract{
Celestial holography   promisingly  reformulates the scattering amplitude holographically in terms of celestial conformal field theory living at null infinity.  Recently, an  infinite-dimensional symmetry algebra  was discovered in Einstein-Yang-Mills theory. The starting point in the derivation is the celestial OPE of  two soft currents, and the key ingredient is the summation of $\overline{SL(2,\mathbb R)}$ descendants in OPE.  In this paper, we consider the supersymmetric Einstein-Yang-Mills theory and obtain the supersymmetric extension of the    holographic symmetry algebra. Furthermore, we derive   infinitely many Ward identities associated with the infinite  soft currents which generate the  holographic symmetry algebra. This is realized by considering the OPE between a soft symmetry current and a hard operator, and then summing over its $\overline{SL(2,\mathbb R)}$ descendants. These Ward identities reproduce the known Ward identities corresponding to the leading, sub-leading, and sub-sub-leading soft graviton theorems as well as the leading and sub-leading soft gluon theorems. By performing   shadow transformations, we also obtain infinitely many shadow Ward identities, including the stress tensor Ward identities for sub-leading soft graviton.  Finally, we   use our procedure to discuss the corrections to Ward identities in effective field theory (EFT), and   reproduce the corrections to  soft theorems at sub-sub-leading order for  graviton and sub-leading order for   photon.  For this aim, we derive   general formulae for the   celestial OPE and its corresponding Ward identities arising from a cubic interaction  of three   spinning massless particles. Our formalism thus provides a unified framework for understanding the Ward identities in celestial conformal field theory, or equivalently the soft theorems in scattering amplitude. 
 
}

\begin{document} 
\maketitle
\flushbottom
\allowdisplaybreaks

\section{Introduction}

Symmetry is arguably one of the most important guiding principles in physics.
The content and power of symmetry have always been evolving with time.
Recent studies have enriched the content of symmetry from continuous to discrete, from ordinary symmetry to higher form symmetry,  and even from symmetry to  non-invertible symmetry. 
In quantum field theory, symmetry and anomaly constrain  the renormalization group flows severely. 
 While in  the swampland program, criteria regarding symmetry  shed  new light on quantum gravity. 
Undoubtedly, a better understanding of symmetry we have, a deeper aspect of fundamental physics  we can acquire.  In particular, symmetry enables us to establish Ward identities  and   conservation laws. 

In this paper, we will be considering the general relativity (GR) and Yang-Mills   (YM) theory  as well  their supersymmetrizations.   Being well understood theoretically and precisely tested experimentally, GR and YM theory  are the cornerstones of modern physics.
In spite, the   symmetry aspect remains not fully clear. For example, it was discovered more than half a century ago that   asymptotic flat spacetime admits,  besides the  Poincare symmetries, infinite dimensional BMS symmetries~\cite{Bondi:1962px,Sachs:1962wk}.  However, the important role of BMS symmetry was never fully appreciated until less than a decade ago due to the pioneering work \cite{Strominger:2013jfa,He:2014laa}, where the Weinberg’s soft graviton theorem was reinterpreted as the Ward identity of BMS symmetry.   Since then, many interesting relations  among soft theorem, memory effect and  asymptotic symmetry  are established. See \cite{Strominger:2017zoo} for a review. 
Furthermore, after many years of studies, a new holographic approach, called celestial holography, to studying quantum gravity in flat spacetime   starts to emerge \cite{Cheung:2016iub,Pasterski:2016qvg}. According to the dictionary of celestial holography,  the scattering particle in the  bulk spacetime can be represented as an  operator  $\cO_{\Delta, J}  $ in celestial conformal field theory (CCFT) living at the boundary null infinity, and  the Mellin transformed scattering amplitudes, called celestial amplitudes,  are just given by the conformal correlators of these  celestial operators \cite{Pasterski:2016qvg,Pasterski:2017kqt}. See \cite{Raclariu:2021zjz,Pasterski:2021rjz} for introductions to  celestial holography.

Celestial holography seems to be a very natural  language in revealing the underlying hidden symmetries of quantum theories. 
In this  celestial approach,   each non-trivial symmetry in the bulk spacetime corresponds to some  soft current in CCFT. An important virtue is that this approach is free from  ambiguities associated with gauge choices, boundary counter-terms, and falloff conditions, which are  sometimes   subtle and tedious in the direct bulk approach \cite{Guevara:2021abz}. 
Another virtue is that this also makes  the computations of algebra efficient   due to mature techniques in 2D CFT. 
 In particular,  the soft theorems of the scattering amplitudes, which are consequences of asymptotic symmetries, are just equivalent to the Ward identities of soft currents in 2D CCFT. The soft currents    are special types of celestial operators $\cO_{\Delta, J}$ with   conformal dimension $\Delta=1, 0, -1, \cdots$ for bosonic fields and $\Delta=1/2, -1/2, -3/2, \cdots $ for fermionic fields. Operators with these special values of conformal dimension will be called soft operators, while all the rest  are hard operators. 
 So far, the Ward identities have been established at several leading orders, $\Delta=1,0,-1$ for graviton and $\Delta=1,0$ for gluon. One main goal in this paper is to derive the Ward identities associated with all the soft currents, namely for all  values of $\Delta$  listed above.

%

Recently, an infinite   tower of symmetries  was discovered  for Einstein-Yang-Mills (EYM)  theory  in celestial holography \cite{Guevara:2021abz,Strominger:2021lvk}. \footnote{~See also \cite{Banerjee:2020zlg, Banerjee:2020vnt,Pasterski:2021fjn} for related discussions. } Each symmetry is associated with a  holomorphic current, which arises from the $\overline{\text{SL}(2,\mathbb R)}$ decomposition of the soft current,  and their commutators are also determined. As such, there seems to be an asymmetry between holomorphic and anti-holomorphic parts  which deserves   explanations. For each celestial operator and soft current, its spin  $J$ in 2D, or equivalently the  helicity in 4D, can be either positive or negative. Both positive and negative helicity soft particles generate  some symmetry. However, whenever two soft particles with opposite helicity are scattered, the resulting amplitude is ambiguous and depends on the order of taking soft limits \cite{He:2015zea}. This can also be observed from the    operator product expansion (OPE) of celestial operators, and ambiguities indeed arise  if two operators with opposite spin are both taken soft. Therefore, there seems to be some intrinsic incompatibility between positive and negative helicity soft operators.  
To sidestep this subtlety, the authors in \cite{Guevara:2021abz} focus on  positive helicity soft
currents only and consider CCFT   with only   $\text{Vir}\otimes \overline{\text{SL}(2,\mathbb R)}$ symmetry, which is the subgroup of the superrotation    $\text{Vir}\otimes \overline{\text{Vir}}$.~\footnote{~Note that we will also work in the (2,2) signature of spacetime and treat $z,\bar z$ as independent   variables (except in section \ref{ShadowWardId} and appendix \ref{integral}).}
With $\overline{\text{SL}(2,\mathbb R)}$  global symmetry, one can decompose every positive helicity soft  current into various chiral currents.  To determine the commutators of these chiral currents, \cite{Guevara:2021abz} considers the OPE of two celestial operators, which arises from the collinear limit of scattering amplitude.  The important ingredient in their derivation is that one also needs to sum over all the $\overline{{\text{SL}(2,\mathbb R)}}$ descendants in OPEs.  Taking soft limits for both   positive helicity operators, which is unambiguous in this case,   and decomposing the resulting soft currents into chiral currents, they obtain the desired algebra of these infinitely many chiral currents. Although such a holographic symmetry algebra only involves positive helicity soft particles, 
it  is infinite dimensional and  thus   an interesting  subalgebra of the full symmetry algebra  in gravity and gauge theory, which remains to uncover.
To emphasize the role  of these chiral currents, we will also refer to this symmetry algebra as \emph{holographic chiral algebra}.

In this paper, we will consider the supersymmetric Einstein-Yang-Mills theory and  derive the   corresponding holographic symmetry algebra  by summing over the $\overline{{\text{SL}(2,\mathbb R)}}$ descendants  of two soft currents. The resulting algebra is thus the supersymmetric extension of that in  \cite{Guevara:2021abz}. 
As in the case of EYM theory,  we will see that the whole algebra is actually generated by the several leading order soft currents by successive commutators.  As such, the new symmetries actually do not impose new extra constraints on S-matrix.  This is expected from  the celestial OPE as the starting point of this derivation. Indeed as shown in \cite{Pate:2019lpp}, one can bootstrap the celestial OPEs by  conformal invariance as well as the soft theorems at the  several leading orders.

With these infinitely many symmetries, it is then natural to ask how do  these symmetries act on celestial operators and celestial amplitudes in CCFT? Furthermore, what are the corresponding Ward identities associated with these  infinite symmetries? In   this paper, we will show that the answers to  these questions   turn out to be simple: instead of considering the OPEs of two soft currents, we just need to start with the OPE between a soft current and a hard operator. Then  after further summing over the $\overline{{\text{SL}(2,\mathbb R)}}$ descendants similarly, we can obtain a resummed  OPE between a soft current and a hard operator, which enables us to read off the action of  soft current on the hard operator and establish the corresponding Ward identities. Using this approach, we establish infinitely many Ward identities for the infinite number of symmetries.  In particular, we  reproduce the known Ward identities corresponding to the leading, sub-leading, and sub-sub-leading soft graviton theorems as well as the leading and sub-leading soft gluon theorems.   The  Ward identities up to these leading orders generate all the rest of Ward identities associated with the rest of symmetries.  The same method applies for fermionic  symmetries associated to soft gluino and soft gravitino. We give the very explicit formulae for all the Ward identities.  Importantly, we must emphasize that our Ward identities  are also applicable to hard operators with negative helicities, as the celestial OPE between a soft current and a hard operator is always unambiguous, regardless of the helicities.

Furthermore, we perform the shadow transformation on the Ward identities. This gives rise to infinitely many shadow Ward identities associated with the infinite shadow symmetries. 
The shadow transformation seems to  play an important role in celestial holography. In particular, the shadow transformation of  sub-leading soft graviton current just gives the stress tensor of 2D celestial CFT, and the   shadow transformation of  Ward identity associated to   sub-leading soft graviton   is just the standard  stress tensor Ward identity in 2D CFT. Our infinite shadow Ward identities are thus a straightforward generalization of this idea to all the rest of symmetries. But it remains to understand what kind  of Ward identity is the most natural one in celestial holography.

 Last but not least, we also make some attempts to    understand   how robust the holographic chiral algebra  is and whether  there are corrections to  Ward identities.  We will try to  address this question in the framework of effective field theory   (EFT) by inspecting the role of various higher dimensional  effective field theory operators. The    EFT approach to studying    corrections to soft theorems was already adopted in \cite{Elvang:2016qvq}.  As in  \cite{Elvang:2016qvq}, we will consider the cubic interactions which involve  three massless particles with arbitrary helicities.  We will first derive a general formula for celestial OPE arising from such a cubic vertex. This is possible as the three-point on-shell amplitude is uniquely fixed by their helicities, up to the coupling constant. With this celestial OPE, we   repeat our procedure as before  and obtain a general formula for Ward identities \eqref{generalWdId} and its shadow cousin \eqref{generalSdWdId}.
 Applying the general result to EFT, we especially reproduce the corrections to  soft theorems at sub-sub-leading order for  graviton and   sub-leading order for   photon found in  \cite{Elvang:2016qvq}. While for the holographic chiral algebra itself, we show that it is robust and free from corrections in EFT, on condition that we   consider the case with only positive helicity soft particles  where our formalism applies.  Beyond this range of applicability, the fate of  holographic chiral algebra is unknown, and a full understanding of soft negative helicity particles   is required.

This paper is organized as follows. In section~\ref{preliminary} we introduce some important preliminaries, including the  celestial OPEs in supersymmetric EYM theory, the $\overline{SL(2,\mathbb R)}$-descendant   summation formula and the    mode   decompositions of soft currents. 
In section~\ref{softalg}, we consider the soft-soft OPEs and derive the holographic chiral algebra in supersymmetric EYM theory. 
In section~\ref{wardId},  we consider the soft-hard  OPEs and derive infinite many Ward identities. 
In section~\ref{ShadowWardId}, we derive the shadow Ward identities.
In section~\ref{EFT}, we apply our formalism in EFT and discuss corrections to the algebra and  Ward identities. 
Finally,   we conclude in section~\ref{conclusion} and discuss some open questions for future research. 
This paper also includes four appendices. 
In appendix~\ref{superOPEs}, we rewrite the celestial  OPEs in terms of celestial on-shell superfields in a compact way. 
In appendix~\ref{integral}, we discuss some integrals which are vital for shadow transformation. 
In appendix~\ref{opeEFT}, we derive a general formula for celestial OPE arising from a cubic interaction of three spinning massless particles. 
In appendix~\ref{softphoton}, we review the soft photon theorems with magnetic corrections. 

 \vspace{.4cm}
 \noindent{\bf Note added. } While we were completing the paper, we learned that the new paper \cite{Himwich:2021dau}  also obtained the celestial OPEs for arbitrary spinning operators,
 which is derived in our  appendix~\ref{opeEFT}  and  subsection~\ref{genope11}.

\section{Preliminaries}\label{preliminary} 

In this section, we will collect some key techniques which are very useful in the rest of the paper.  As the starting point, we will first review the celestial OPEs \eqref{ope1}-\eqref{ope10} in supersymmetric EYM theory.  As we discussed in the introduction, the key ingredient   is to sum over all the $\overline{SL(2,\mathbb R)}$  descendant contributions in the OPE. Therefore for later convenience,  we will then present a general formula  \eqref{OPEdes} realizing this goal. Finally, we will discuss some aspects of the soft symmetry currents and particularly its   decomposition into various chiral currents \eqref{Rk} under   $\overline{SL(2,\mathbb R)}$. 

\subsection{OPEs in supersymmetric EYM theory}
Let us first review the  celestial OPEs  in $\cN=1$ supersymmetric EYM theory. The OPEs can be obtained from the Mellin transformation of collinear limits of scattering amplitudes \cite{Fan:2019emx,Fotopoulos:2020bqj,Jiang:2021xzy}. Alternatively, they can be bootstrapped using    conformal invariance and soft theorems \cite{Pate:2019lpp}. 
 Especially we will be mainly focusing on the operators corresponding to positive helicity particles which are not ambiguous in the soft limit. Their explicit OPEs read:~\footnote{~We use the convention: $f^{abc}{}_\text{here}=-if^{ab}{}_c \; {}_{\text{\cite{Pate:2019lpp,Guevara:2021abz}}  }$ and $\kappa_{\text{\cite{Pate:2019lpp,Guevara:2021abz}}}=2_\text{ here}$. Also for simplicity we will assume that all the particles are out-going throughout this paper.   }
\beqn\label{ope1}
\cO^a_{\Delta_1, +1}(z_1, \bar z_1) \cO^b_{\Delta_2, +1}(z_2, \bar z_2) 
&\sim&  \frac{f^{abc} }{z_{12}} B(\Delta_1-1, \Delta_2-1) \cO^c_{\Delta_1+\Delta_2-1, +1}(z_2, \bar z_2) ~,
\\
\cO _{\Delta_1, +2}(z_1, \bar z_1)\cO _{\Delta_2, +2}(z_2, \bar z_2) &\sim&
- \frac{\bar z_{12}}{z_{12}}  B(\Delta_1-1, \Delta_2-1) \cO_{\Delta_1+\Delta_2, +2}(z_2,\bar z_2)  ~,
\\
\cO^a_{\Delta_1, +1}(z_1, \bar z_1) \cO _{\Delta_2, +2}(z_2, \bar z_2) &\sim&
- \frac{\bar z_{12}}{z_{12}}  B(\Delta_1 , \Delta_2-1 ) \cO^a_{\Delta_1+\Delta_2, +1}(z_2,\bar z_2)  ~,
\\
\cO^a_{\Delta_1, +\frac12}(z_1, \bar z_1) \cO^b_{\Delta_2, +\frac12}(z_2, \bar z_2)  &\sim& 0~,
\\
\cO^a_{\Delta_1, +\frac12}(z_1, \bar z_1) \cO _{\Delta_2, +\frac32}(z_2, \bar z_2)  &\sim& 0~,
\\
\cO _{\Delta_1, +\frac32}(z_1, \bar z_1) \cO _{\Delta_2, +\frac32}(z_2, \bar z_2)  &\sim& 0~,
\\
\cO^a_{\Delta_1, +\frac12}(z_1, \bar z_1) \cO^b_{\Delta_2, +1}(z_2, \bar z_2) 
  &\sim&
  \frac{f^{abc} }{z_{12}} B(\Delta_1-\frac12, \Delta_2-1) \cO^c_{\Delta_1+\Delta_2-1, +\frac12}(z_2, \bar z_2)~,
\\
\cO^a _{\Delta_1, +\frac12}(z_1, \bar z_1) \cO _{\Delta_2, +2}(z_2, \bar z_2) 
&\sim& - \frac{\bar z_{12}}{z_{12}} B(\Delta_1  +\frac12, \Delta_2-1    ) \cO^a_{\Delta_1+\Delta_2 , +\frac12}(z_2, \bar z_2) ~,
\\
\cO^a_{\Delta_1,+1}(z_1, \bar z_1) \cO_{\Delta_2, +\frac32} (z_2, \bar z_2) &\sim&
-\frac{\bar z_{12}}{z_{12}}  B(\Delta_1 , \Delta_2-\frac12)  \cO^a_{\Delta_1+\Delta_2, +\frac12} (z_2, \bar z_2)~,
\label{ope9}
\\
\cO_{\Delta_1,+\frac 32} (z_1, \bar z_1) \cO_{\Delta_2,+2} (z_2, \bar z_2)  &\sim&
-\frac{\bar z_{12}}{z_{12}}  B(\Delta_1-\frac12, \Delta_2-1) \cO_{\Delta_1+\Delta_2, +\frac32}(z_2,\bar z_2)  ~,
\label{ope10}
\eeqn
where $z_{12}=z_1-z_2$ and in all the OPEs we only keep the leading singular terms, if they exist. It is worth explaining the notation a bit here.  We use  $\cO_{\Delta, J} $ to denote the celestial  operator with dimension $\Delta$ and spin $J$ in celestial CFT. Note that $J$ also coincides with the helicity of particles in 4D bulk spacetime. We also have the standard CFT relation  $\Delta=h+\bar h, J=h-\bar h$ where $h$ and $\bar h$ are holomorphic and anti-holomorphic conformal weights, respectively.  So $\cO_{\Delta,+2}$,  $\cO_{\Delta,+3/2}$ correspond to celestial graviton and gravitino operators, while $\cO^a_{\Delta,+1}$,  $\cO^a_{\Delta,+1/2}$ are   celestial gluon and gluino operators. Here $a$ is the color index of the gauge group and $f^{abc}$ is the corresponding anti-symmetric  structure constant.

These OPEs are consistent with supersymmetry.  As shown in appendix~\ref{superOPEs}, one can make supersymmetry manifest by introducing    on-shell celestial superfields for each multiplet and  then write down the corresponding super-OPEs. The supersymmetry transformation rules can be obtained easily. In particular,  the  supersymmetry  acts on the gravity multiplet as ($\alpha,\dot\alpha=1,2$)
\beqn\label{Qsusy}
&&
Q^\alpha\cdot \cO_{\Delta,+\frac32}(z,\bz) =z^{\alpha-1} \cO_{\Delta+\frac12, +2}(z,\bz)~, \qquad 
Q^\alpha\cdot \cO_{\Delta,+ 2}(z,\bz) =0~,
\\ &&
\tilde Q^{\dot\alpha}\cdot \cO_{\Delta,+ 2}(z,\bz) =\bz^{\dot\alpha-1} \cO_{\Delta+\frac12, +\frac32}(z,\bz)~, \qquad
 \tilde Q^{\dot\alpha}\cdot \cO_{\Delta,+ \frac32}(z,\bz)=0~,
 \label{Qsusy2}
 \eeqn
and similarly for the vector multiplet.  One can check that the OPEs above indeed transform consistently under the supersymmetry actions \eqref{Qsusy} and \eqref{Qsusy2}.~\footnote{~Since we are considering the leading term in the OPE, it turns out to be sufficient to just consider the case $\alpha,\dot\alpha=1$ as $z_1\approx z_2$ to leading order. Besides, a  useful identity to show SUSY invariance is $B(x,y)=B(x,y+1)+B (x+1,y)$. }

For OPEs involving graviton minimally coupled matter, they actually take the following universal  form with the same couplings due to  the equivalence principle: 
\footnote{~Actually this is valid even for non-positive helicity particle  with $J_1\le 0$ except that there may be  also an extra piece which is singular in the limit $\bz_{12}\to 0$. }
\be\label{gravityOPE}
\cO_{\Delta_1,  J_1} (z_1, \bar z_1)   \cO _{\Delta_2, +2}(z_2, \bar z_2)  \sim 
- \frac{\bar z_{12}}{z_{12}}  B(\Delta_1-J_1+1, \Delta_2-1 ) \cO_{\Delta_1+\Delta_2,  J_1}(z_2,\bar z_2)~ , \qquad J_1>0 ~.
\ee
 Applying the supersymmetric transformation to \eqref{gravityOPE}, we get the general formula of OPE involving gravitino:  \footnote{Again, one can consider the case $J_1\le 0$ but with more complications. See \cite{Fotopoulos:2020bqj} for explicit formulae. }
\be\label{gravitinoOPE}
\cO_{\Delta_1,  J_1} (z_1, \bar z_1)   \cO _{\Delta_2, +\frac32}(z_2, \bar z_2)  \sim 
- \frac{\bar z_{12}}{z_{12}}  B(\Delta_1-J_1+1, \Delta_2-\frac12 ) \cO_{\Delta_1+\Delta_2,  J_1-\frac12}(z_2,\bar z_2)~,  \qquad J_1\in \mathbb Z_+~,
\ee
where  $\cO_{\Delta_1,  +J_1} $ should be a bosonic operator,   otherwise the OPE is regular in supersymmetric EYM theory. And $ \cO_{\Delta_1' ,  J_1-\frac12}$ is the supersymmetric partner of $\cO_{\Delta_1 ,  J_1} $.

Before closing this subsection, let us quote some very useful formulae for Beta and Gamma functions which   come from the following integral: 
 \be\label{betafcn}
B(x,y) =\int_0^1 t^{x-1} (1-t)^{y-1} dt=\frac{\Gamma(x)\Gamma(y)}{\Gamma(x+y)}~.
\ee
In particular, $B(x,y)$ and $\Gamma(x)$ have simple poles at non-positive integral argument. More specifically, we have
\beqn
\lim_{x\to k} \Gamma(x +m) &=&\frac{1}{x-k} \frac{(-1)^{-m-k}}{(-m-k)!}~, \qquad\qquad\qquad\qquad k+m= 0,-1,\cdots  ~,
\\ \label{Betapole2}
\lim_{x\to k } B(x+m,y )&=&
\frac{1}{x-k} \frac{(-1)^{-m-k}}{(-m-k)!}\frac{\Gamma(y)}{\Gamma(k+m+y)}~,
\qquad \quad k+m= 0,-1,\cdots ~\neq y, 
\\
\label{Betapole}
\lim_{x\to k,y\to l} B(x+m,y+l)&=&
\frac{x+y-k-l}{(x-k)(y-l)} \PBK{-m-n-k-l \\  -m-k }~, 
 \qquad k+m, \; l+n= 0,-1,\cdots~, \qquad\qquad
\eeqn
where the binomial $\PBK{n\\m}=\frac{n!}{m!(n-m)!}$ and $0!=1$.

%
%

\subsection{Summing over $\overline{SL(2,\mathbb R)}$ descendants}
One of the key ingredient in this paper is the summation of descendants in OPE which leads to OPE block. 
Generally    the OPE of two primary operators is given by
\be\label{OPE1}
  \cO_{\Delta_1, J_1}(z_1,\bar z_1)\cO_{\Delta_2,J_2}(z_2,\bar z_2)    \sim  \sum_{O_P}     C_{\cO_1\cO_2}^{\cO_P}  
  \frac{\cO_{\Delta_P, J_P}(z_2, \bz_2)}{(z_{12} \bz_{12})^{\frac{\Delta_1+\Delta_2-\Delta_P} {2} }(z_{12} /\bz_{12})^{\frac{J_1+J_2-J_P}{2}}}+\cdots~,
\ee
where $\cO_{\Delta_P, J_P}$  are  primary operators and dots represent all the descendants.  We would like to   include  the contributions of all the 
$\overline{SL(2,\mathbb R)}$  descendants. This is can be nicely realized by  replacing each primary operator (together with the kinematic factors) with its corresponding OPE block.  More specifically, the
$\overline{SL(2,\mathbb R)}$ OPE block is given by \cite{Czech:2016xec}
\beqn
 \overline{SL(2,\mathbb R)} \text{ OPE block}
 & =&\int_{\bz_2}^{\bz_1} {d\bz_3\;  \cO_{ \bar h_P }( \bar z_3 })
 \EV{\cO_{ \bar h_1 }( \bar z_1)\cO_{ \bar h_2 }( \bar z_2) \tilde \cO_{1- \bar h_P }( \bar z_3 )}
  \\& =&
  \int_{\bz_2}^{\bz_1}\frac{d\bz_3\; \cO_{ \bar h_P }( \bar z_3 )} {\bz_{12}^{\bar h_1+\bar h_2+\bar h_P-1}
 \bz_{32}^{\bar h_2-\bar h_1-\bar h_P+1} \bar z_{13}^{\bar h_1-\bar h_2-\bar h_P+1}}~,
\eeqn
where $ \tilde \cO_{1- \bar h_P }=\widetilde{\cO_{ \bar h_P}}$ is the shadow of $\cO_{\bar h_P}$ and has weight  $1-\bar h_P$. 

After  summing over the $\overline{SL(2,\mathbb R)}$  descendants in \eqref{OPE1}, we then arrive at \cite{Guevara:2021abz}
   \beqn\label{OPEdes}
 && \cO_{\Delta_1, J_1}(z_1,\bar z_1)\cO_{\Delta_2,J_2}(z_2,\bar z_2)    
\nonumber   \\ &\sim&
     \sum_{O_P}
  \mathcal N_{\cO_1\cO_2}^{\cO_P}   \frac{\bz_{12}^{N-M}}{z_{12}^{M+N}} 
\int_0^1  dt \; \cO_{ \Delta_P,J_P } (z_2, \bar z_2+t \bar z_{12}) 
\; t^{\Delta_1 -J_1-M+N -1 }(1-t)^{ \Delta_2 -J_2-M+N-1 }~,
\qquad
 \eeqn
 where
  \be
 M=\frac{\Delta_1+\Delta_2-\Delta_P}{2}~, \qquad N=\frac{J_1+J_2-J_P}{2}~,
 \ee
 and the coefficient $  \mathcal N_{\cO_1\cO_2}^{\cO_P}   $ can be fixed by comparing the leading term in \eqref{OPEdes} with \eqref{OPE1}: 
 \be
  \mathcal N_{\cO_1\cO_2}^{\cO_P}  =  \frac{C_{\cO_1\cO_2}^{\cO_P}     }{B(\Delta_1 -J_1-M+N,\;\Delta_2 -J_2-M+N)}~.
 \ee 

This formula \eqref{OPEdes} is the key ingredient in the rest of paper. 
 
\subsection{Mode decomposition of soft currents  }

Following the dictionary of celestial holography, the  Mellin transformed scattering amplitude  can be regarded as the correlator of celestial operators $\cO_{\Delta, J}$ in CCFT.  In order to form a complete basis,  the dimension  should reside in the principal continuous series of the
unitary representations of  $SL(2,
\mathbb C)$: $\Delta\in 1 +i \mathbb R$ \cite{Pasterski:2017kqt}. However, we can also analytic continue $\Delta$ in the complex plane.   In particular, for special  values of $\Delta$, they actually generate large gauge transformations at null infinity, and are thus the symmetry generators. Operators with these special values of dimension are called soft  operators, while the rest of are called hard operators.

 More specifically, for positive spin-$J$ operator,  the  soft symmetry currents  are  defined as ~\footnote{~Actually, we   should exclude $k=2$ for graviton and $k=3/2$ for gravitino as soft currents.
 As we will see, they are the central terms in the algebra,   and do not act on hard operators. So
 the honest  dimension of    soft currents takes values in  $k=1, 0, -1, \cdots $ for bosonic fields, and $k=1/2, -1/2, -3/2, \cdots$ for fermionic fields.  } 
 \be\label{sfotmdoe}
 R^{k ,J}(z,\bar z)=\lim_{\Delta\to k} (\Delta-k) \cO_{\Delta, +J}(z,\bz), \qquad k=J, J-1, J-2, \cdots~.
 \ee
 It has weights $(h, \bar h) =(\frac{k+J}{2}, \frac{k-J}{2})$.
 
These soft symmetry currents admit mode expansions under $\overline{SL(2,\mathbb R)}$~\cite{Guevara:2021abz}:
\beqn\label{Rk}
R^{k ,J}(z,\bar z)
&=&\sum_{n=\frac{k-J}{2}}^{\frac{ J-k}{2}} \frac{R_n^{k,J }(z)}{\bar z^{n+\frac{k-J}{2}}}
=\bar z^{J-k}R_{\frac{k-J}{2}}^{k,J }(z)+\bar z^{ J-k-1}R_{\frac{k-J+2}{2}}^{k,J }(z)+\cdots +R_{\frac{J-k}{2}}^{k,J }(z)~.
\eeqn
This gives rise to $J-k+1$ holomorphic  currents $R^{k,J}_n(z)$ which will be referred to as \emph{chiral   currents}. They all have the same holomorphic weight $h=(k+J)/2$ and transform in the ($J-k+1$)-dimensional representational of $\overline{SL(2,\mathbb R)}$. 

As we will see, it turns out to be more convenient to rescale each mode and redefine the chiral   currents as follows \cite{Strominger:2021lvk}: 
\be\label{modeRedefine}
\mathcal R_n^{i,J} = (i-1-n)! (i-1+n)! R_n^{J+2-2i,J}~,
\ee
where
\be \label{eqin}
 n=1-i,\,  2-i, \cdots i-1, \quad
i =\frac{J-k}{2}+1=1, \, \frac32,\,  2, \cdots~.
\ee

Physically, \eqref{modeRedefine} corresponds to a light-transformation \cite{Strominger:2021lvk}. In general, the light transformations     along two null directions for operator with weights $(h,\bar h)$ in 2D CFT are given by \cite{Kravchuk:2018htv}
 \be
   {\mathbf L}[\cO](w,\bar z)=\int d  z\; (  w-  z)^{2  h-2} \cO(z,\bz)~, \qquad\quad
  \bar {\mathbf L}[\cO](z,\bar w)=\int d\bar z\; (\bar w-\bar z)^{2\bar h-2} \cO(z,\bz)~.
 \ee
 Applying the second  light transformation to our soft operator yields 
 \footnote{~Here we use the following formula to evaluate the integral
  \be\nonumber
 \int_{-\infty}^\infty d\bar z\;  (\bar w-\bz)^{-a} \bar z^{-b} 
 =-  {2\pi  {\rm i} }  \frac{  \bar w^{-a-b+1}\Gamma( a+b-1) }{\Gamma(a)\Gamma(b)}
 =  -  {2   {\rm i} } \sin(\pi b) \frac{  \bar w^{-a-b+1}\Gamma( a+b-1) \Gamma(1-b)}{\Gamma(a) }~,
  \ee
where we evaluate the integral using    Mathematica for $\Real a<1,\; \Real b<1,\; \Real (a+b)>1$ and then perform analytic continuation.  
In the second equality, we use the identity  $\Gamma(x) \Gamma(1-x)=\frac{\pi}{\sin(\pi x)}$.
  }
   \beqn
&& \epsilon \bar {\mathbf L}[ \cO_{k+\epsilon, J}](z,\bar w)
=  \bar {\mathbf L}[ R^{k ,J}](z,\bar w)
\\& =&
\sum_{n=\frac{k-J}{2}}^{\frac{ J-k}{2}}  R_n^{k,J }(z) 
\int d\bar z\; (\bar w-\bar z)^{k+\epsilon-J-2}  \bar z^{-n+\frac{J-k-\epsilon}{2}} 
\\&=&
\sum_{n=\frac{k-J}{2}}^{\frac{ J-k}{2}}  R_n^{k,J }(z)   
(-2\pi  {\rm i} )  w^{-1-n+\frac{k+\epsilon-J}{2}} \frac{ \Gamma(n+1-\frac{k+\epsilon-J}{2}) \Gamma(1-n-\frac{k+\epsilon-J}{2})}
{\Gamma( 2+J-k-\epsilon) }\frac{\sin \Big( \pi (n+\frac{k+\epsilon-J}{2})\Big) }{\pi}~. 
\qquad \qquad
\label{ltsf}
 \eeqn
Note in the above equations, we need to deform the dimension $k$ by $\epsilon$ before applying the light transformation, but finally we need to take the limit $\epsilon \to 0$. It is easy to see that in \eqref{ltsf} we can   just  set $\epsilon $ to 0 everywhere except for $\sin \Big( \pi( n+\frac{k+\epsilon-J}{2})\Big)=(-)^{n+(k-J)/2}\sin \frac{\pi\epsilon}{2} \approx  \frac{\pi \epsilon}{2} (-)^{n+\frac{k-J}{2}} $.  This  $\epsilon$ just cancels with $\epsilon=\Delta-k$ in the definition  of soft currents. After a  change of  variable using \eqref{eqin},  the equation \eqref{ltsf}  gets simplified  
   \beqn
   \bar {\mathbf L}[ \cO_{J+2-2i, J}](z,\bar z)
 &=& -  {\rm i}  \pi
\sum_{n= 1-i}^{i-1 }       (-)^{n+1-i}
  \bar z^{-i-n} \frac{ \Gamma(i+n) \Gamma(i-n)} {\Gamma( 2i) }  \;  R_n^{J+2-2i,J }(z)   
\\&=&  \pi  {\rm i}  \frac{  (-)^{2i} }{{\Gamma( 2i) } }
\sum_{n= 1-i}^{i-1 }    
  \frac{  \cR^{i,J}_n}{(-\bar z)^{ i +n}}~,
     \eeqn
 where $  \cR_n^{i,J}$ is precisely the same as that defined in \eqref{modeRedefine}.  Therefore \eqref{modeRedefine} is indeed equivalent to a light transformation, and $  \cR_n^{i,J}$ is exactly the mode expansion of light-transformed soft operator, up to a mode-dependent sign and an overall constant. 
 \footnote{ Using this relation, one can infer the OPEs between   hard operators and light-transformed soft operators 
 from the OPEs between   hard operators and  chiral currents, which we will compute in section~\ref{wardId}. 
 Alternatively, \cite{Himwich:2021dau}  directly computed the OPEs between hard operators and  light-transformed soft operators.}
 
In order to   extract each chiral currents $R^k_n(z)$ from soft symmetry currents $R^k(z,\bz)$ in \eqref{Rk}, we can take derivatives for multiple times: 
  \be\label{derpR0}
 \bar\p^pR^{k,J }(z,\bar z)=\sum_{n=\frac{k-J}{2}}^{\frac{ J-k}{2}} ( -n-\frac{k-J}{2}-p+1)_p \bar z^{-n-\frac{k-J}{2}-p} {R_n^{k,J }(z)}{ }
 =p!  R_{-\frac{k-J}{2}-p} ^{k,J }(z) +\sO(\bz)~,
 \ee
 where $\sO(\bz)$ are terms which have anti-holomorphic dependence on $\bz$. Therefore, by considering holomorphic terms on the left hand side, we unambiguously  select the specific chiral current $R_{-\frac{k-J}{2}-p}^{k,J }(z)$.

For redefined chiral currents in \eqref{modeRedefine}, we similarly have 
  \be\label{derpR}
 \bar\p^pR^{k,J }(z,\bar z) 
  =p!  R_{-\frac{k-J}{2}-p}^{k,J }(z) +\sO(\bz)
  =\frac{1}{(i-1+n)!}\mathcal R_n^{i,J} (z)+\sO(\bz)~,
 \ee
where 
\be
k=J+2-2i~, \qquad p=i-1-n~, \qquad n=\frac{J-k}{2}-p~, \qquad  i=\frac{J-k}{2}+1~.
\ee

Finally, we will use   different symbols $H,I, K,L $ to label  the soft currents for graviton, gravitino, gluon and gluino. As a consequence, we have the following notation: 
 \beqn
&&  H^k=  R ^{k, +2}~, \qquad
 I^k=  R^{k, +3/2 }~, \qquad
 K^{k,a}=   R ^{k, +1, a}~, \qquad
 L^{k,a}=   R^{k, +1/2, a}~, \qquad
\\
&& 
\label{modenotation}
 \mathcal H^{i}_n=  \cR_n^{i, +2}~, \qquad
 \mathcal I^{i}_n=  \cR_n^{i, +3/2 }~, \qquad
  \mathcal K^{i, a}_n=  \cR_n^{i, +1, a}~, \qquad
 \mathcal L^{i, a}_n=  \cR_n^{i, +1/2, a}~. \qquad
 \eeqn

\section{Holographic chiral algebra from soft-soft OPEs}\label{softalg}

In this section, we will derive the holographic symmetry algebra  in supersymmetric EYM theory. This is realized by considering the celestial OPEs  \eqref{ope1}-\eqref{ope10} where both operators are taken soft. After summing over all the $\overline{SL(2,\mathbb R)}$ descendants using \eqref{OPEdes} and decomposing each soft current into chiral currents with \eqref{Rk}, we arrive at the   OPEs of chiral currents. This just yields the holographic symmetry algebra, or more precisely the holographic chiral algebra,  of the chiral currents. Such an  infinite-dimensional algebra is thus the underlying  hidden symmetry of scattering amplitude.
As we will see,  these symmetries are not all independent. Instead, they are generated by   several leading soft currents.  
Our discussion  in this section is the supersymmetric generalization of \cite{Guevara:2021abz}.

We will first discuss the holographic chiral algebra    in the pure SYM case  which  contains only gluons and gluinos. Then we will include gravitons and gravitinos and derive the full   holographic chiral algebra.

\subsection{Pure SYM theory}
Let us  first discuss the pure SYM theory involving gluons and gluinos only.  In such a case, the OPE can be generally written as
  \be \label{YMJ12}
\cO^a_{\Delta_1, +J_1}(z_1, \bar z_1) \cO^b_{\Delta_2, +J_2}(z_2, \bar z_2) 
\sim  \frac{f^{abc} }{z_{12}} B(\Delta_1-J_1 , \Delta_2-J_2) 
  \cO^c_{\Delta_1+\Delta_2-1, +(J_1+J_2-1)}(z_2, \bar z_2) ~,
\ee
where $J_1, J_2=1,\frac12$ and $ J_1+J_2=2,\frac32$. Note that the OPE between two gluino operators is regular. 

As described many times before, the key ingredient  here is to sum over all the $\overline{SL(2,\mathbb R)}$ descendants in the OPE \cite{Guevara:2021abz}.  Using the general formula \eqref{OPEdes}, we get
   \beqn\label{gluonOPE}
 \cO^a_{\Delta_1, J_1}(z_1,\bar z_1)\cO^b_{\Delta_2,J_2}(z_2,\bar z_2) &\sim& 
  \frac{f^{abc} }{z_{12}} 
\int_0^1  {dt \;  \cO^c_{\Delta_1+\Delta_2-1, +(J_1+J_2-1)}(z_2, \bar z_2+t \bar z_{12})} \;
{t^{\Delta_1 -J_1-1 }(1-t)^{ \Delta_2 -J_2 -1 }}   ~.
\qquad\quad
 \eeqn
 
 Since the goal in this section is to  derive the algebra of   chiral currents, we thus need to separate different  contributions in \eqref{Rk} for each soft symmetry  current.    This can be realized by taking derivatives with respect to the anti-holomorphic coordinates.  Applying such derivatives to \eqref{gluonOPE} yields
 \beqn \label{OJOJope}
&&  \bar\p^p \cO^a_{\Delta_1, J_1}(z_1,\bar z_1)   \bar\p^q \cO^b_{\Delta_2,J_2}(z_2,\bar z_2)  
  \\& \sim&
  \frac{f^{abc} }{z_{12}} 
\int_0^1  {dt \;   \bar\p^{p+q} \cO^c_{\Delta_1+\Delta_2-1, +(J_1+J_2-1)}(z_2, \bar z_2+t \bar z_{12})} \;
{t^{\Delta_1 -J_1-1+p }(1-t)^{ \Delta_2 -J_2 -1+q }}   
\qquad\qquad\qquad \nonumber
\\& \sim&
  \frac{f^{abc} }{z_{12}} 
  \; \sum_{s=0}^\infty  \frac{  \bar z_{12}^s}{s!} \bar\p^{p+q+s} \cO^c_{ \Delta_1+\Delta_2-1,+(J_1+J_2-1)} (z_2, \bar z_2 ) 
B(\Delta_1 -J_1 +p+s ,\Delta_2 -J_2 +q)~.
\label{opegluon2}
 \eeqn
 
 To further discuss the case of soft symmetry currents defined in \eqref{sfotmdoe}, we  just need to focus on     special values of conformal dimension $\Delta_1\to k,\Delta_2\to l$ where $k-J_1,l-J_2=0, -1, -2,\cdots$.
As one can see from \eqref{derpR}, the leading  term in $  \bar\p^p O$ is a  purely holomorphic   current. Therefore,  in order to extract the contribution from   chiral currents in \eqref{OJOJope},
  we just need to keep   $s=0$ term in \eqref{opegluon2} which is independent of anti-holomorphic coordinates $\bz_1,\bz_2 $. As a consequence, we find 
 
  \be
  \bar\p^p R^{k,J_1,a} (z_1,\bar z_1)   \bar\p^q R^{l,J_2,b}(z_2,\bar z_2)  \sim
  \frac{f^{abc} }{z_{12}} 
  \PBK{-k-l+J_1 +J_2-p-q \\  -k+J_1  -p }
  \bar\p^{p+q } R ^{ k+l-1,J_1+J_2-1,c } (z_2, \bar z_2 ) +\sO( \bz_1,\bz_2)~,
 \ee 
 where we used the residue  formula \eqref{Betapole}.  Further using \eqref{derpR0}, we get
  \be\label{RRope}
   R_{\frac{J_1-k}{2}-p}^{k,J_1,a} (z_1 )    R_{\frac{J_2-l}{2}-q}^{l,J_2,b}(z_2 )  \sim
  \frac{f^{abc} }{z_{12}} 
  \PBK{-k-l+J_1 +J_2-p-q \\  -k+J_1  -p }
\PBK{p+q \\ p } R_{\frac{J_1+J_2-k-l}{2}-p-q} ^{ k+l-1,J_1+J_2-1,c } (z_2  ) ~.
 \ee 
 
Alternatively, we can use the rescaled chiral currents defined in \eqref{modeRedefine}, which simplifies  \eqref{RRope} to
  \be
  \cR_n^{i,J_1,a} (z_1 )    \cR_m^{j,J_2,b}(z_2 )  \sim
  \frac{f^{abc} }{z_{12}} 
\cR_{n+m} ^{ i+j-1,J_1+J_2-1,c } (z_2 ) ~.
 \ee 
 Such a dramatic simplification was one of the motivation for introducing the  rescaled chiral currents in \eqref{modeRedefine}. 

Spelling out the case $J_1,J_2=1,1/2$ explicitly and using the notation \eqref{modenotation}, we finally obtain \beqn\label{gluonOPE2}
 \mathcal K^{i,a}_n(z) \mathcal K^{j,b}_m(0) &\sim& \frac{f^{abc}}{z} \mathcal K^{i+j-1,c}_{n+m}(0)  ~,\label{KK}\\
  \mathcal K^{i,a}_n(z) \mathcal L^{j,b}_m(0) &\sim& \frac{f^{abc}}{z} \mathcal L^{i+j-1,c}_{n+m}(0) ~,
  \label{KL}\\
    \mathcal L^{i,a}_n(z) \mathcal L^{j,b}_m(0) &\sim& 0~. \label{LL}
 \eeqn

 \subsection{Supersymmetric EYM  theory}
 
 Now we want to include gravity. As we show in \eqref{gravityOPE}, there is a universal OPE between graviton operator and matter operator which are minimally coupled.  
 Just like the case of  the SYM theory discussed above, we need to  use  \eqref{OPEdes} to sum over   $\overline{SL(2,\mathbb R)}$ descendants in the OPE \eqref{gravityOPE}. This then gives
\be\label{opeGraviton}
\cO_{\Delta_1, J_1} (z_1, \bar z_1)   \cO _{\Delta_2, +2}(z_2, \bar z_2) \sim
- \frac{\bar z_{12}}{z_{12}} 
\int_0^1  dt \; \cO_{ \Delta_1+\Delta_2,J_1 } (z_2, \bar z_2+t \bar z_{12}) 
\; t^{\Delta_1 -J_1  }(1-t)^{ \Delta_2 -2   }~.
\ee
Taking derivatives yields
\beqn
&&
\bar\p^p \cO_{\Delta_1, J_1} (z_1, \bar z_1)  \bar\p^q \cO _{\Delta_2, +2}(z_2, \bar z_2) 
\\
  &\sim & 
- \frac{\bar z_{12}}{z_{12}}  
  {   \sum_{s=0}^\infty   \frac{(\bar z_{12})^s}{s!} \bar\p^{p+q+s} \cO_{\Delta_1+\Delta_2, J_1}(z_2,\bar z_2 )} \;
B(\Delta_1 -J_1 +p +s+1, \Delta_2 -1+q )
\nonumber \\  && 
- \frac{p}{z_{12}}  
  {   \sum_{s=0}^\infty   \frac{(\bar z_{12})^s}{s!} \bar\p^{p+q+s-1} \cO_{\Delta_1+\Delta_2, J_1}(z_2,\bar z_2 )} \;
B(\Delta_1 -J_1 +p +s , \Delta_2 -1+q )
 \label{gope1} \\  && 
+ \frac{q}{z_{12}}  
  {   \sum_{s=0}^\infty   \frac{(\bar z_{12})^s}{s!} \bar\p^{p+q+s-1} \cO_{\Delta_1+\Delta_2, J_1}(z_2,\bar z_2 )} \;
B(\Delta_1 -J_1 +p +s+1 , \Delta_2 -2+q )~.
\label{gope2}
\qquad 
\eeqn

Next we want to find the OPE between  soft symmetry currents defined in \eqref{sfotmdoe}. Just as in the case of SYM, we need to take the limit    $\Delta_1\to k, \Delta_2\to l$, and only keep the $s=0$ terms  in \eqref{gope1} and \eqref{gope2} which are  independent of $\bz_1 $ and $\bz_2$.    
As a result, we arrive at
\beqn \label{gOPE3}
&&
\bar\p^p R^{k, J_1} (z_1, \bar z_1)  \bar\p^q R^  {l, +2}(z_2, \bar z_2)
\\& \sim&
- \frac{1 }{z_{12}}  
 \frac{(-k -l+J_1  -p-q +1)! }{(-k +J_1 -p )! (-l+2-q)! }\Big(  p(2-l-q)  -q(J _1 - k-p) \Big)
  {    \bar\p^{p+q-1} R^{k+l, J_1}(z_2,\bar z_2 )} 
+\sO(\bz_1,\bz_2 )~,
\nonumber
\eeqn
 where we used the    equation  \eqref{Betapole} and the following formula to simplify the result
\beqn
&&
\Bigg[ p \PBK{-k -l+J_1  -p-q   +1 \\-k +J_1 -p }  -q  \PBK{-k -l+J_1  -p-q   +1 \\-k +J_1 -p-1 }  \Bigg]
\nonumber\\&=& \frac{(-k -l+J_1  -p-q +1)! }{(-k +J_1 -p )! (-l+2-q)! }\Big(  p(2-l-q)  -q(J _1 - k-p) \Big)~.
\eeqn

Using~\eqref{derpR}, the OPE between (redefined) chiral currents then can be straightforwardly obtained:
  \be
\cR^{i,J_1}_n(z_1) \cR^{ j,+ 2}_m (z_2)   \sim  - \frac{2 }{z_{12}}   \Big(   m(i-1) -n(j-1) \Big)  
 \cR^{i+j-2,J_1}_{n+m}(z_2)  ~.
\ee
 
 Writing out the above formula explicitly for $J_1=2,3/2,1,1/2 $ with notation  \eqref{modenotation}, we get the OPE between chiral graviton current  and the chiral current for graviton, gravitino, gluon and gluino:  
\beqn\label{HHope}
\cH^i_n (z) \cH^j_m (0)  &\sim&  - \frac{2 }{z}   \Big(   m(i-1) -n(j-1) \Big)  
 \cH^{i+j-2 }_{n+m}(0)  ~,\label{HH}
 \\
 \cI^i_n (z) \cH^j_m (0) &\sim&   - \frac{2 }{z}   \Big(   m(i-1) -n(j-1) \Big)  
 \cI^{i+j-2 }_{n+m}(0)   ~,\label{IH}
  \\
 \cK^{i,a}_n (z) \cH^j_m (0) &\sim&   - \frac{2 }{z}   \Big(   m(i-1) -n(j-1) \Big)  
 \cK^{i+j-2,a }_{n+m}(0)   ~, \label{KH}
  \\
 \cL_n^{i,a} (z) \cH^j_m (0) &\sim&   - \frac{2 }{z}   \Big(   m(i-1) -n(j-1) \Big)  
 \cL^{i+j-2,a }_{n+m}(0)   ~.\label{LH}
\eeqn

The final non-trivial OPE we need to consider is the one between gluon and gravitino \eqref{ope9}.  
Instead of focusing on  this special OPE, we can again consider the OPE between gravitino and arbitrary bosonic operator, \footnote{~Note that the OPE between two fermionic operators is regular. } which is described in the general formula \eqref{gravitinoOPE}. We can then repeat the same steps for SYM and graviton to find the OPE between chiral currents. Without spelling out any details, we just write down the final result:  
\be
\cR^{i,J_1}_n(z_1) \cR^{ j,+ \frac32}_m (z_2)   \sim  - \frac{2 }{z_{12}}   \Big(   m(i-1) -n(j-1) \Big)  
  \cR^{i+j-2, J_1-\frac12 }_{n+m}(z_2)  ~.
\ee
For graviton $J_1=2$, this agrees with \eqref{IH}.  And the OPE between chiral gluon current and chiral gravitino current is given by:
\beqn
  \cI^i_n (z) \cK^{j,a}_m (0) &\sim&   - \frac{2 }{z}   \Big(   m(i-1) -n(j-1) \Big)  
 \cL^{i+j-2,a }_{n+m}(0) ~.  \label{IK}  
  \eeqn
 Finally we have two fermionic OPEs which are regular:
 \beqn
   \cI^i_n (z) \cI^{j }_m (0) &\sim&  0~, \qquad \label{II} \\
     \cL^i_n (z) \cI^{j }_m (0) &\sim&  0~. \qquad \label{LI} 
    \eeqn
 
Now we obtain all the OPEs between chiral currents  \eqref{KK},\eqref{KL},\eqref{LL},\eqref{HH}, \eqref{IH}, \eqref{KH}, \eqref{LH}, \eqref{IK},\eqref{II}, \eqref{LI}.
 They all have at most simple poles. These OPEs can be rewritten as commutators by employing the following formula \cite{DiFrancesco:1997nk}: 
 \be
[A,B](z) =\oint_z \frac{dw}{2\pi i} A(w) B(z) ~.
\ee 
Applying this formula to all the OPEs, we get the commutators between all the chiral fields.   This gives rise to the holographic symmetry algebra \cite{Guevara:2021abz}. Since all the symmetries are generated by chiral currents, we will also refer to such algebra as \emph{holographic chiral algebra}. 
In particular, applying the formula to  graviton case \eqref{HHope}, we get the commutator between chiral graviton currents  \be
[\cH^i_n , \; \cH^j_m    ]=  - 2 \Big(   m(i-1) -n(j-1) \Big)  
 \cH^{i+j-2 }_{n+m} ~. 
\ee
This turns out be just the $w_{1+\infty}$ algebra as observed in  \cite{Strominger:2021lvk}.~\footnote{~To compare with \cite{Strominger:2021lvk}, we need to redefine $  w_n^i=\frac12 \cH^i_n$. }
Note that $\cH^1_0$ commutes with all  $\cH^i_n$ and is thus a central term. Also it is easy to see that $ \cH^{3/2}_n,\cH^{2}_n,\cH^{5/2}_n$ generate all the rest of $\cH^i_n$ by successive commutators. Physically, this means that the algebra is actually generated by the leading, sub-leading, and sub-sub-leading soft gravitons.  Similarly, in the case of gluon, the  corresponding algebra is generated by the leading and sub-leading soft gluons.  These are not surprising, as we start with OPEs which can actually be bootstrapped from these soft theorems  \cite{Pate:2019lpp}.

The commutators for the rest of chiral currents can be obtained similarly in an obvious way as all OPEs involved   have only simple poles. Furthermore, we can also consider Einstein-Maxwell theory and its supersymmetrization. The resulting algebra is almost identical except that we need to remove the color index and set $f^{abc}$ to zero because Maxwell theory  has no self-interaction.

\subsection{Structures in holography chiral  algebra}
 \paragraph{Supersymmetry.} So far, we obtained the OPEs between all chiral currents. Since we are considering the supersymmetric  theory,  these OPEs should be invariant under the supersymmetry transformations. 
Indeed, applying \eqref{Qsusy}   \eqref{Qsusy2} to \eqref{Rk}  \eqref{modeRedefine},  one can easily derive the following supersymmetry transformation rules acting on chiral currents: 
\beqn
\cQ^\alpha\cdot \cR_n^{i, J}(z) &=& 
z^{\alpha-1}  \cR_n^{i,J+\frac12}(z)~,\qquad\qquad\qquad\qquad\quad
\cQ^\alpha\cdot \cR_n^{i, J+\frac12}(z) =0~, \quad J=\frac32, \frac12~, 
\\
\tilde\cQ^{\dot\alpha}\cdot \cR_n^{i, J}(z) &=&
\Big(i-1-n(2\dot\alpha-3) \Big)  \cR_{n-\frac32+\dot\alpha}^{ i-\frac12,J-\frac12}(z)~,\qquad
\tilde\cQ^{\dot\alpha}\cdot \cR_n^{i, J+\frac12}(z) =0~, \quad J=2,1~.\qquad\qquad
\eeqn
Then one can explicitly check that all the chiral OPEs and thus the holographic chiral algebra   are indeed invariant under these SUSY transformation rules.

\paragraph{Generalized Sugawara construction. } In the   OPE of two  chiral gluon current \eqref{KK}, we see that the structure is very similar to the Kac-Moody algebra at level zero, if we only keep the color indices. 
Furthermore, given a Kac-Moody algebra, one can then naturally construct the Sugawara stress tensor.

More specifically, we can consider the leading soft current $ \cK^{1,a}_0 $, which is also a chiral current. It has weights $(h,\bar h)=(1,0)$ and the OPE between $ \cK^{1,a}_0 $ with itself just gives the  Kac-Moody algebra at level zero. Therefore we can naturally the  following Sugawara stress tensor: 
\be
\cT(z)=\gamma :\cK^{1,a}_0\cK^{1,a}_0:(z)\equiv\gamma\oint_z \frac{dw}{2\pi i} \frac{1}{w-z}\cK^{1,a}_0(w)\cK^{1,a}_0(z)~,
\ee
where the sum over color index $a$ should be understood, $\gamma$ is a constant to be fixed below and  $: \cdots : $ is the normal order product that we defined above. 
Then we find 
\beqn\label{OPETK}
\cT(z) \cK^{i, a}_n(0)   \sim   \frac{ \cK^{i, a}_n(0)}{z^2}+\frac{\p \cK^{i, a}_n(0)}{z}~, 
 \qquad
 \cT(z) \cT(0) \sim \frac{2\cT(0)}{z^2}+\frac{\p\cT(0)}{z}~,
\label{TTope}
\eeqn
if we choose   $\gamma$ such that
\be
-\gamma f^{abc} f^{bcd}=  \delta^{ad}~.
\ee
Hence $\gamma$ is inversely proportional to the dual Coxeter number of gauge group. Therefore $\cT$ indeed behaves as a stress tensor in this soft sector. This Sugawara construction was considered before in \cite{Fan:2020xjj}.
\footnote{~In \cite{Fan:2020xjj}, they also discussed OPE between $\cT$ and hard operator, and found that $\cT$ constructed in this way does not behave properly anymore. }

 More generally, we also  find
 \footnote{~Here we need to use the property  $:\cK^{i, a}_n\cK^{j, b}_m: =: \cK^{j, a}_m \cK^{i, b}_n:$
which follows from the mode expansion of  $:\cO^a_{\Delta_1, +1}\cO^b_{\Delta_2, +1}:=:\cO^a_{\Delta_2, +1}\cO^b_{\Delta_1, +1}:$, which  is an operator with dimension $\Delta_1+\Delta_2$.
Similarly, we also have $:\cK^{i, a}_n\cL^{j, b}_m: =: \cK^{j, a}_m \cL^{i, b}_n:$. An extra identity is needed in the fermionic case:  
$  \p\cL^{i+j-1,c}_{n+m} =-2\gamma f^{abc}:\cK^{i, a}_n\cL^{j, b}_m: $, 
which is just a null state relation. 
}
  \beqn
 \gamma :\cK^{j_1,b}_{m_1}\cK^{j_2,b}_{m_2}:(z)\cK^{i, a }_n(0) &\sim&  \frac{ \cK^{i+j_1+j_2-2, a}_{n+m_1+m_2} (0)}{z^2}+\frac{\p\cK^{i+j_1+j_2-2, a}_{n+m_1+m_2} (0) }{z}~,
\\
 \gamma :\cK^{j_1,b}_{m_1}\cK^{j_2,b}_{m_2}:(z)\cL^{i, a }_n(0) &\sim&  \frac{ \cL^{i+j_1+j_2-2, a}_{n+m_1+m_2} (0)}{z^2}+\frac{\p\cL^{i+j_1+j_2-2, a}_{n+m_1+m_2} (0) }{z}~.
 \eeqn
 So it also holds if we replace $\cK $ with $\cL$ in  \eqref{OPETK}. This is the generalized Sugawara construction whose physical   implications remain to be understood.

%
%
%
%
\section{Ward identities from soft-hard OPEs}  \label{wardId}

In the last section, we   obtained the holographic chiral algebra  in supersymmetric EYM theory. They are the hidden symmetry of scattering amplitude. A natural question to ask is how do these symmetries act on amplitude? And what are the physical consequences of these symmetries?

In this section, we will show that the infinite number of soft symmetry currents just lead to an infinite number of Ward identities. Each Ward identity just relates celestial amplitudes with and without the insertion of soft current. The way to derive these Ward identities is almost identical to the derivation of symmetry algebra in the last section, except that we need to use  the soft-hard OPE now. 
More specifically, we will  pick the celestial OPEs  \eqref{ope1}-\eqref{ope10} and take one of them to be soft.  After summing over all the $\overline{SL(2,\mathbb R)}$ descendants, we will arrive at the OPE between soft currents and hard operators. It turns out that this just yields the Ward identity of the corresponding soft current. If we further decompose the soft current into chiral currents using \eqref{Rk}, we arrive at the chiral Ward identities for the chiral currents.

These Ward identities reproduce the known Ward identities corresponding to the leading, sub-leading and sub-sub-leading soft graviton theorems as well as the leading and sub-leading soft gluon theorems. Since the whole tower of soft currents is generated by these several leading soft currents, the resulting tower of Ward identities is thus   also generated by the Ward identities of these several leading order soft currents.  Nevertheless, our formulae for all the Ward identities are explicit and may shed new light on the structure of holographic chiral algebra. This method also works for fermionic symmetry current, but we will not discuss their  corresponding Ward identities  explicitly in this section. Instead,  we will present a general OPE formula  \eqref{opeeft} from which a general formula for Ward identities \eqref{generalWdId} is derived. 
Specializing the general formula   \eqref{generalWdId}  to soft gluino or soft gravitino,  one easily  obtains their corresponding Ward identities.
 
 \subsection{Graviton Ward identity}\label{secGdWd}
  As we discussed, our basic strategy is to  consider the OPE with all  $\overline{SL(2,\mathbb R)}$  descendants included, and then specialize to the case of   soft-hard OPE. 
  Let us first discuss the case of graviton. The general OPE involving graviton after summing over $\overline{SL(2,\mathbb R)}$  descendants  is given in    \eqref{opeGraviton}: 
  \be \label{Ogg2}
\cO_{\Delta_1, +2} (z_1, \bar z_1)   \cO _{\Delta_2,   J_2}(z_2, \bar z_2) \sim
- \frac{\bar z_{12}}{z_{12}} 
\int_0^1  dt \; \cO_{ \Delta_1+\Delta_2, J_2 } (z_2, \bar z_2+t \bar z_{12}) 
\; t^{\Delta_1 -2   }(1-t)^{ \Delta_2 -J_2   }~.
\ee

After performing the integral on the right hand side, we get
 \beqn
 &&\cO_{\Delta_1, +2 } (z_1, \bar z_1)   \cO _{\Delta_2, J_2}(z_2, \bar z_2) 
 \\
&\sim&
- \frac{\bar z_{12}}{z_{12}}  \sum_{s=0}^\infty
   {  \; \frac{(\bar z_{12})^s}{s!} \p^s \cO_{\Delta_1+\Delta_2, +J_2}(z_2,\bar z_2 )} \;
 B(\Delta_1  +s-1 , \Delta_2-J_2+1)~.
\eeqn

We need to consider the soft graviton for $\cO_{\Delta_1, +2 }   $, so we set  $\Delta_1\to k$.   Then the above OPE simplified as follows:
 \beqn\label{HOhard}
&&H^k(z_1, \bar z_1)   \cO _{\Delta_2, J_2}(z_2, \bar z_2) 
\nonumber \\&\sim&
- \frac{\bar z_{12}}{z_{12}}  \sum_{s=0}^{1-k}
   {  \; \frac{(\bar z_{12})^s}{s!} \bar\p^s \cO_{ \Delta_2+k, +J_2}(z_2,\bar z_2 )} \;
 \frac{(-1)^{-s+1-k}}{(-s+1-k)!}    \frac{ \Gamma( \Delta_2-J_2+1)}{\Gamma( \Delta_2-J_2+k+s )}~,
 \eeqn
where we used  the formula \eqref{Betapole2} and the infinite sum of $s$ truncates because we need a pole from $\Gamma(\Delta_1+s-1)$.  This is the OPE between the soft graviton current and a hard operator.  

We also want to find the OPE between chiral currents and the hard operators.   For this purpose, we expand $(\bz_1-\bz_2)^{s+1}$:
 \beqn\label{HOope}
  &&H^k(z_1, \bar z_1)   \cO _{\Delta_2, J_2}(z_2, \bar z_2) 
\nonumber  \\&\sim&
- \frac{1}{z_{12}}   \sum_{s=0}^ {1-k} \sum_{n=0}^{s+1} \PBK{s+1\\n } 
\frac{ \bar z_1^n (-\bar z_2)^{s+1-n}}{s!} 
\bar \p^s \cO_{ \Delta_2+k, +J_2}(z_2,\bar z_2 ) 
 \frac{(-1)^{-s+1-k}}{(-s+1-k)!}    \frac{ \Gamma( \Delta_2-J_2+1)}{\Gamma( \Delta_2-J_2+k+s )}
 \qquad\qquad \\&\sim&
- \frac{1}{z_{12}}   \sum_{n=0}^{2-k}  \sum_{s=\max(0,n-1)}^ {1-k } \PBK{s+1\\n } 
\bar z_1^n   \bar z_2^{s+1-n}
\bar \p^s \cO_{ \Delta_2+k, +J_2}(z_2,\bar z_2 ) 
 \frac{(-1)^{-n-k}}{s!( -s+1-k )!}    \frac{ \Gamma( \Delta_2-J_2+1)}{\Gamma( \Delta_2-J_2+k+s )}~.\quad
  \qquad
 \eeqn
 Note  the non-trivial exchange of sums over $s$ and $n$. 
On the other hand, the soft current can also be expanded in terms of chiral currents using \eqref{Rk} and \eqref{modeRedefine}. Inserting the expansion into \eqref{HOope} and comparing the left and right hand sides, we find the following OPE 
  \beqn\label{HOchiral}
 \cH^i_n(z_1 )   \cO _{\Delta_2, J_2}(z_2, \bar z_2) 
 &\sim&
  \frac{  (-1)^{n  + i  }}{z_{12}}     \sum_{r=\max(0,n-i+2)}^ {n+i-1 }  
  \PBK{i+n-1 \\r }  
      \frac{(r-n+i-1)  \Gamma( \Delta_2-J_2+1)}{\Gamma( \Delta_2-J_2  +r-n- i+2    )}
\nonumber  \\&&  \qquad\qquad\qquad\qquad\qquad
\times  \bar z_2^r
\;\bar \p^{r-n+i-2} \cO_{\Delta_2-2i+4,  J_2}(z_2,\bar z_2 ) ~.
  \eeqn

Spelling out the OPE explicitly at several leading orders, we get \footnote{Note $2\bar h=\Delta-J$. Also  $\cH^1_0 $ is a central term and  acts on hard operators trivially as one can see from \eqref{HOhard} by setting $k\to 2$. }

$\bullet$  leading soft graviton ($\Delta_H=1 $):
 \beqn
\cH^{3/2}_{1/2} (z_1 )   \cO _{\Delta , , J}(z_2, \bar z_2) \sim &&
\frac{\bar z_2 }{z_{12}} \cO _{\Delta +1, J}(z_2, \bar z_2)~,
\\
\cH^{3/2}_{-1/2} (z_1 )   \cO _{\Delta , J}(z_2, \bar z_2) \sim &&
\frac{ -1}{z_{12}} \cO _{\Delta +1, J}(z_2, \bar z_2)~,
 \eeqn

$\bullet$ sub-leading soft graviton ($\Delta_H=0 $):
 \beqn
  \frac12\cH^{ 2}_{1} (z_1)  \cO _{\Delta , J}(z_2, \bar z_2) \sim &&
- \frac{\bar z_2^2\p_{\bar z_2} +2\bar h \bar z_2  }{z_{12}} \cO _{\Delta   , J}(z_2, \bar z_2), \qquad  \\
\cH^{ 2}_{0}  (z_1 )   \cO _{\Delta , J}(z_2, \bar z_2) \sim &&
2 \frac{   \bar z_2 \p_{\bar z_2}+   \bar h  }{z_{12}} \cO _{\Delta  , J}(z_2, \bar z_2)~,
\\
\frac12 \cH^{ 2}_{-1}  (z_1 )   \cO _{\Delta , J}(z_2, \bar z_2) \sim &&
- \frac{ \p_{\bar z_2}  }{z_{12}} \cO _{\Delta, J}(z_2, \bar z_2)~,
 \eeqn
 
 $\bullet$  sub-sub-leading soft graviton ($\Delta_H=-1 $):
 \beqn
\frac16 \cH^{ \frac52}_{\frac32} (w )   \cO _{\Delta , J}(z, \bar z) \sim &&
  \frac{\bar z ^3\p_\bz^2 +4\bar h \bz^2 \p_\bz +2\bar h(2\bar h-1) \bz  }{2(w-z)} \cO _{\Delta -1, J}(z , \bz)~,  
  \\
\frac12 \cH^{ \frac52}_{\frac12} (w )   \cO _{\Delta , J}(z, \bar z) \sim &&
-  \frac{3\bz ^2\p_\bz^2 +8\bar h \bz \p_\bz +2\bar h(2\bar h-1)    }{2(w-z)} \cO _{\Delta -1, J}(z , \bz)~,  
  \\
\frac12 \cH^{ \frac52}_{-\frac12}(w )   \cO _{\Delta , J}(z, \bar z) \sim &&
  \frac{3\bz  \p_\bz^2 +4\bar h   \p_\bz     }{2(w-z)} \cO _{\Delta -1, J}(z , \bz)~,    
    \\
\frac16\cH^{ \frac52}_{-\frac32} (w )   \cO _{\Delta , J}(z, \bar z) \sim &&
 - \frac{\p_\bz^2  }{2(w-z)} \cO _{\Delta -1, J}(z , \bz)~, \qquad 
 \eeqn
 where the coefficients on the left hand side of OPE is just the rescaling factor in \eqref{modeRedefine}.

 The OPEs \eqref{HOope}, \eqref{HOchiral}  as well as the explicit forms in the several leading orders just give the transformation rules  of hard operators under the action soft symmetry current and chiral current. 
Up to sub-sub-leading order, it  turns out the structures here are identical to  that in  \cite{Puhm:2019zbl,Pate:2019lpp,Banerjee:2020zlg}. 

Using the  OPEs   \eqref{HOope}, \eqref{HOchiral}   in correlators, we claim that we have the following Ward identities for soft symmetry currents
\footnote{\label{caveat}
Note that the  Ward identities in the present form are not complete beyond sub-sub-leading order for $l\le -2$. The holomorphic weight of the graviton current $H^l$ is $h=(l+2)/2$, and hence is non-positive $h\le 0$ for  $l\le -2$. As a result, the graviton current $H^l$ scales as $z^{-2h}$ and does not decay to zero at $\infty $.  This indicates that some polynomial terms in $z$ are missing in the present Ward identities. The polynomial terms here are reminiscent of the   remainder terms appearing  in the soft expansion of graviton amplitude beyond sub-sub-leading order, see e.g. \cite{Li:2018gnc}.  
This kind of polynomial terms can be killed by considering the zero mode of the soft current: one can multiply both sides of \eqref{gWd} with some holomorphic function $f(z)$ and then perform a contour integral along a contour which encloses all the hard operator insertions. The resulting   Ward identity is then complete and exact for all $H^l$, as the polynomials are killed by the contour integral. However, the polynomial terms would play a role if one considers non-zero mode or   the correlation function involving descendants. The polynomial terms seem to play a role for the consistency of $w_{1+\infty}$ algebra. It would be interesting to understand whether the missing polynomial is universal and try to   figure out  them explicitly.   
Similar caveats apply to all the rest of Ward identities in this paper.  We would like to thank Andrew Strominger for comments and in particular Shamik   Banerjee  for useful discussions.  
}
  \beqn\label{gWd}
\nonumber&&\EV{H^l(z , \bar z)   \cO_{\Delta_1,J_1} (z_1,\bar z_1)\cdots \cO_{\Delta_m,J_m} (z_m,\bar z_m)}
 \\&=&\sum_{k=1}^m \sum_{s=0}^{1-l}
  \frac{(\bz-\bz_k)^{s+1}}{z-z_k} 
 \frac{(-1)^{-s -l}}{s!(-s+1-l)!}    \frac{ \Gamma(2\bar h_k+1)}{\Gamma(2\bar h_k +l+s )}
 \nonumber \\&&\qquad\times
     \bar\p_k^s \EV{  \cO_{\Delta_1,J_1} (z_1,\bar z_1)\cdots\cO_{ \Delta_k+l,  J_k}(z_k,\bar z_k )\cdots \cO_{\Delta_m,J_m} (z_m,\bar z_m)}~,
 \eeqn
and furthermore the chiral Ward identities for chiral currents: 
\beqn\label{chiralWd}
&&\EV{ \cH^i_{n}(z ) \cO_{\Delta_1,J_1} (z_1,\bar z_1)\cdots \cO_{\Delta_m,J_m} (z_m,\bar z_m)}
\nonumber \\&=& 
\sum_{k=1}^m  \frac{  (-1)^{n  + i  }}{z-z_k}     \sum_{r=\max(0,n-i+2)}^ {n+i-1 }  
  \PBK{i+n-1 \\r }  
      \frac{(r-n+i-1)  \Gamma( 2\bar h_k+1)}{\Gamma( 2\bar h_k +r-n- i+2    )}
\nonumber  \\&&  \qquad\qquad 
\times  \bar z_k^r
\;\bar \p_k^{r-n+i-2} \EV{\cO_{\Delta_1,J_1} (z_1,\bar z_1)\cdots\cO_{ \Delta_k-2i+4,  J_k}(z_k,\bar z_k ) 
 \cdots \cO_{\Delta_m,J_m} (z_m,\bar z_m)}~,
\eeqn
where $\bar\p_k=\frac{\p}{\p \bar z_k}$.

This is not surprising: in the standard stress tensor Ward identity we essentially also replace every pair of primary operator  and stress tensor with their corresponding singular OPEs. This seems to be a general feature for all Ward identities arising from symmetry in CFT. Since here all the currents also correspond to some symmetries, similar tricks should also work.  Actually, in the present case, these Ward identities can be further justified as follows. Note that  in OPEs \eqref{HOope}, \eqref{HOchiral}, all the  $\overline{SL(2,\mathbb R)}$  descendants have been summed over, implying that the anti-holomorphic dependence is supposed to be exact in the above Ward identities. Therefore, we only  need to worry about the holomorphic dependence because we have not considered the  $SL(2,\mathbb R)$ descendants. However, there is a big simplification in the present case: all the OPEs \eqref{HOope}, \eqref{HOchiral} have only simple poles in the holomorphic coordinates. If we denote the correlator on the left-hand side  of Ward identity as $F(z)$, then we know that such a function can only be singular when $z$ hits other operator insertions $z_i$, and the singular behavior in the coincident limit is dictated by OPEs. Since all the OPEs here have only simple poles, we thus learn that $F(z)$ is a meromorphic function with only simples poles (note the $F(z)$ should be finite at infinity).  Following the Mittag-Leffler   theorem, such a meromorphic function is uniquely determined by its poles and the residues there, up to a constant. Our Ward identities have exactly the expected behavior  as a meromorphic function. \footnote{ Beyond sub-sub-leading order, there may be also  polynomial terms. See footnote \ref{caveat}. }  This thus establishes our Ward identities. 

One can check explicitly that  up to sub-sub-leading order, our Ward identities reproduce all the  known Ward identities corresponding to leading, sub-leading and sub-sub-leading soft theorems~\cite{Adamo:2019ipt,Puhm:2019zbl, Guevara:2019ypd}:
 \beqn    
      &&\EV{H^1(z , \bar z)   \cO_{\Delta_1,J_1} (z_1,\bar z_1)\cdots \cO_{\Delta_m,J_m} (z_m,\bar z_m)}
\nonumber \\&=&  - \sum_{k=1}^m 
  \frac{ \bz-\bz_k }{z-z_k} \EV{  \cO_{\Delta_1,J_1} (z_1,\bar z_1)\cdots\cO_{ \Delta_k+1, +J_k}(z_k,\bar z_k )\cdots \cO_{\Delta_m,J_m} (z_m,\bar z_m)}~,
 \eeqn    
and
  \beqn    
      &&\EV{H^0(z , \bar z)   \cO_{\Delta_1,J_1} (z_1,\bar z_1)\cdots \cO_{\Delta_m,J_m} (z_m,\bar z_m)}
\nonumber \\&=&\sum_{k=1}^m  
  \frac{(\bz-\bz_k)^{2}}{z-z_k} 
 \Big[\frac{2\bar h_k}{\bar z- \bar z_k} - \bar\partial_k\Big] 
  \EV{  \cO_{\Delta_1,J_1} (z_1,\bar z_1)\cdots 
     \cO_{\Delta_m,J_m} (z_m,\bar z_m)}~,
 \eeqn    
 and
 \beqn    
      &&\EV{H^{-1}(z , \bar z)   \cO_{\Delta_1,J_1} (z_1,\bar z_1)\cdots \cO_{\Delta_m,J_m} (z_m,\bar z_m)}
\nonumber \\&=& -\frac12
 \sum_{k=1}^m 
  \frac{(\bz-\bz_k)^{3}}{z-z_k} 
\Big[  \frac {2\bar h_k(2\bar h_k-1)}{(\bz-\bz_k)^2} -\frac {4\bar h_k \bar\p_k}{ \bz-\bz_k } +\bar\p_k^2  \Big]
\nonumber \nonumber \\&&\qquad\times
 \EV{  \cO_{\Delta_1,J_1} (z_1,\bar z_1)\cdots\cO_{ \Delta_k-1,  J_k}(z_k,\bar z_k )\cdots \cO_{\Delta_m,J_m} (z_m,\bar z_m)}~. 
 \eeqn    
The rest of Ward identities are supposed to be guaranteed by the associativity of the holographic chiral algebra that we discussed in the previous section.

One important remark is that our Ward identities \eqref{gWd}\eqref{chiralWd} hold  for both positive and negative helicity  \emph{hard} operators, although we only considered the  positive helicity soft operators   in the discussion of holographic chiral algebra in the previous section.
For example, the   OPE  of two gravitons with opposite helicity is  \cite{Pate:2019lpp}
\beqn
\cO _{\Delta_1, +2}(z_1, \bar z_1)\cO _{\Delta_2, -2}(z_2, \bar z_2) &  \sim &
- \frac{\bar z_{12}}{z_{12}}  B(\Delta_1-1, \Delta_2+3) \cO_{\Delta_1+\Delta_2, -2}(z_2,\bar z_2)  
\nonumber\\&&
 - \frac{ z_{12}}{\bar z_{12}}  B(\Delta_1+3, \Delta_2-1) \cO_{\Delta_1+\Delta_2, +2}(z_2,\bar z_2)  ~.
\eeqn
The first term has been considered in \eqref{Ogg2}, so we only need to worry about the second term.  
The second term has a  zero, instead of a  pole, in the holomorphic coincident limit $z_1\to z_2$. 
Even at sub-sub-leading order $\Delta_1\to k=1,0,-1$, the second term  has no contribution in soft-hard OPE because  $(\Delta_1-k) B(\Delta_1+3, \Delta_2-1)$  vanishes at these orders. As a result, the Ward identities can not be modified up to these orders. Since we know soft gravitons up to sub-sub-leading order generate the whole tower of symmetry, the second term is supposed to have no effect  either for the rest of Ward identities.

\subsection{Gluon Ward identity}
Now we switch to the gluon case. We start with the following gluon OPE \eqref{gluonOPE} where  all the  $\overline{SL(2,\mathbb R)}$  descendants have been included:
\be
\cO_{\Delta_1, +1}^{a}(z_1,\bar z_1) \cO_{\Delta_2, J_2}^b (z_2,\bar z_2)\sim  \frac{f^{abc}}{z_{12} } \int_0^1 dt\;   \cO_{\Delta_1+\Delta_2-1,J_2} (z_2, \bar z_2+t\bar z_{12}) \; t^{ \Delta_1-2}(1-t)^{ \Delta_2-J_2-1}~.
\ee
 Doing the integral on the right hand side gives:
 \beqn
\cO_{\Delta_1, +1}^{a}(z_1,\bar z_1) \cO_{\Delta_2,J_2 }^b (z_2,\bar z_2)
&\sim&
  \frac{f^{abc}}{z_{12} }  \sum_{s=0}^\infty \frac{(\bar z_{12})^s}{s!} \bar\p^s \cO_{\Delta_1+\Delta_2-1,J_2} (z_2, \bar z_2 ) \;B( \Delta_1+s-1, \Delta_2 -J_2)~.
  \qquad\qquad
\eeqn
%
%
%
We then take  $\cO_{\Delta_1,+1}^a$ soft by setting $\Delta_1\to k$. Then the above OPE reduces to
\be
K^{k,a}(z_1,\bar z_1) \cO_{\Delta_2,J_2}^b (z_2,\bar z_2)\sim  \frac{f^{abc}}{z_{12} }  \sum_{s=0}^{1-k} \frac{(\bar z_{12})^s}{s!} \bar\p^s \cO_{ \Delta_2+k-1, J_2} (z_2, \bar z_2 ) \;
 \frac{(-1)^{-s+1-k}}{(-s+1-k)!}    \frac{ \Gamma( \Delta_2-J_2)}{\Gamma( \Delta_2-J_2+k+s-1)}~,
 \ee
where we used \eqref{Betapole2} and the infinite sum of $s$ truncates because we need a pole from $\Gamma(\Delta_1+s-1)$. 
 
To find the OPE between chiral gluon currents and the hard operators, we expand $(\bz_1-\bz_2)^{s  }$:
\beqn
&&
K^{k,a}(z_1,\bar z_1) \cO_{\Delta_2,J_2}^b (z_2,\bar z_2)
\qquad\\&\sim &
  \frac{f^{abc}}{z_{12} }  \sum_{s=0}^{1-k} 
\sum_{n=0}^s \PBK{s\\n }
\frac{ \bar z_1^n (-\bar z_2)^{s-n}}{s!} \bar\p^s  \cO_{ \Delta_2+k-1,J_2 } (z_2, \bar z_2 ) \;
 \frac{(-1)^{-s+1-k}}{(-s+1-k)!}     \frac{ \Gamma( \Delta_2-J_2)}{\Gamma( \Delta_2-J_2+k+s-1)}
 \qquad
\qquad\\
&\sim &
  \frac{f^{abc}}{z_{12} }
 \sum_{n=0}^{1-k}  \sum_{s=n }^ {1-k }  \PBK{s\\n }
\frac{ \bar z_1^n (-\bar z_2)^{s-n}}{s!} \bar\p^s  \cO_{ \Delta_2+k-1,J_2} (z_2, \bar z_2 ) \;
 \frac{(-1)^{-s+1-k}}{(-s+1-k)!}    \frac{ \Gamma( \Delta_2-J_2)}{\Gamma( \Delta_2-J_2+k+s-1)} ~, \qquad\qquad
 \label{OPEKO}
 \eeqn
where we exchange the sum of $s$ and $n$. 

Further inserting the mode expansion into the above OPE and comparing the left and right hand sides, we find the following OPE  between chiral current and hard operator
\beqn
&&
 \cK^{i,a}_{n}(z_1)\cO_{\Delta_2, J_2}^b (z_2,\bar z_2)
\nonumber\\ &\sim&
  \frac{f^{abc}}{z_{12} }
 \sum_{r=0 }^ {i-1+n }     (-1)^{ i-1+n}  \PBK{i-1+n \\ r}
\bar z_2 ^r \bar\p^{r+i-1-n}  \cO_{  \Delta_2+2-2i,J_2} (z_2, \bar z_2 ) \;
   \frac{ \Gamma( \Delta_2-J_2)}{\Gamma( \Delta_2-J_2+1-i-n+r)}~.
   \qquad
\qquad
 \eeqn
 
At first two leading orders, the OPE explicitly reads:

$\bullet$ leading soft gluon ($\Delta_K=1$):

\be
\cK^{1,a}_0(z)\cO_{\Delta ,J}^b (w,\bar w)\sim    \frac{f^{abc}}{z-w }     \cO_{\Delta,J}^c (w,\bar w)~,
\ee

$\bullet$ sub-leading soft gluon ($\Delta_K=0$):
\beqn
\cK^{\frac32,a}_{\frac12}(z)\cO_{\Delta ,J}^b (w,\bar w) &\sim&  - \frac{f^{abc}}{z-w } \Big( (2\bar h-1) +\bar w \bar\p \Big) 
 \cO_{\Delta-1,J}^c (w,\bar w)~,
\\
\cK^{\frac32,a}_{-\frac12}(z)\cO_{\Delta ,J}^b (w,\bar w)&\sim&    \frac{f^{abc}}{z-w } \bar\p   \cO_{\Delta-1,J}^c (w,\bar w)~.
\eeqn
The structure here are again identical to  that in \cite{He:2015zea,Pate:2019mfs,Pate:2019lpp,Banerjee:2020vnt}.
%
%
%
 
As in the graviton case, we can now propose the following Ward identities for soft gluon currents
\beqn\label{gluonWds}
&&\EV{ K^{l,a}(z,\bz) \cO_{\Delta_1,J_1}^{b_1} (z_1,\bar z_1)\cdots \cO_{\Delta_m,J_m}^{b_m} (z_m,\bar z_m)}
\nonumber\\&=& 
\sum_{k=1}^m  f^{ab_k c_k}  \sum_{s=0}^{1-l}
  \frac{ (\bar z-\bz_k)^s  }{z-z_k }    \frac{(-1)^{-s+1-l}}{s!(-s+1-l)!}    \frac{ \Gamma( 2\bar h_k)}{\Gamma(2\bar h_k+l+s-1)}
 \nonumber \\& & 
  \times
  \bar\p_k^s
     \EV{\cO_{\Delta_1,J_1}^{b_1} (z_1,\bar z_1)  \cdots  \cO^{c_k}_{ \Delta_k+l-1, J_k}  (z_k, \bar z_k )  \cdots \cO_{\Delta_m,J_m}^{b_m}  (z_1,\bar z_m)}~,
\eeqn
and the chiral Ward identities for chiral gluon currents
\beqn\label{chilragluonWd}
&&\EV{ \cK^{i,a}_{n}(z ) \cO_{\Delta_1,J_1}^{b_1} (z_1,\bar z_1)\cdots \cO_{\Delta_m,J_m}^{b_m} (z_m,\bar z_m)}
\nonumber\\&=& 
\sum_{k=1}^m \frac{f^{ab_k c_k}}{z-z_k }
 \sum_{r=0 }^ {i-1+n }     (-1)^{ i-1+n}  \PBK{i-1+n \\ r}
   \frac{ \Gamma( 2\bar h_k)}{\Gamma( 2\bar h_k+1-i-n+r)}
\nonumber  \\& & 
  \times
    \bar z_k ^r \bar\p_k^{r+i-1-n}   
     \EV{\cO_{\Delta_1,J_1}^{b_1} (z_1,\bar z_1)  \cdots  \cO^{c_k}_{ \Delta_k+l-1, J_k}  (z_k, \bar z_k )  \cdots \cO_{\Delta_m,J_m}^{b_m}  (z_1,\bar z_m)}~,
\eeqn
where the hard operators in the vector multiplet have helicity  $J_k=\pm \frac12, \pm 1$. \footnote{~More generally, a similar type of Ward identity is supposed to hold even for hard operators belonging to  the chiral multiplets which have helicities $J=0, \pm 1/2$ and  transform in some representation $\mathbf R$ of the gauge group, but one needs to replace $f^{abc}$ with some representation matrix $(T_{\mathbf R}^a)_{IJ} $. }
The arguments for the validity of these Ward identities are similar to that in the graviton case we discussed before.  And it is easy to check that they are in perfect agreement with the Ward identities for leading and sub-leading soft gluon theorems \cite{Fan:2019emx,Nandan:2019jas,Pate:2019mfs} :
\beqn
&&
\EV{  {K^{1,a}}(z,\bar z)   \cO_{\Delta_1,J_1}^{b_1} (z_1,\bar z_1)\cdots \cO_{\Delta_m,J_m}^{b_m} (z_m,\bar z_m)}
 \nonumber\\&=&  
\sum_{k=1}^m  
\frac{ f^{ab_k c_k}  }{  z- z_k}      \EV{\cO_{\Delta_1,J_1}^{b_1} (z_1,\bar z_1)  \cdots  \cO^{c_k}_{ \Delta_k , J_k}  (z_k, \bar z_k )  \cdots \cO_{\Delta_m,J_m}^{b_m}  (z_m,\bar z_m)}~,\qquad\quad
 \eeqn
and
\beqn\label{subphoton}
&&
\EV{  {K^{0,a}}(z,\bar z)   \cO_{\Delta_1,J_1}^{b_1} (z_1,\bar z_1)\cdots \cO_{\Delta_m,J_m}^{b_m} (z_m,\bar z_m)}
\\&=&  
\sum_{k=1}^m  
\frac{f^{ab_k c_k}  }{  z- z_k}    \Big[ -({2\bar h_k -1})  + (\bar z-\bz_k)  {\bar\p_k} \Big]     \EV{\cO_{\Delta_1,J_1}^{b_1} (z_1,\bar z_1)  \cdots  \cO^{c_k}_{ \Delta_k -1, J_k}  (z_k, \bar z_k )  \cdots \cO_{\Delta_m,J_m}^{b_m}  (z_m,\bar z_m)}~. 
\nonumber
 \eeqn

 \subsection{Photon Ward identity}

 For Maxwell-matter coupled system, like QED, the scattering amplitudes also   factorize when the photon is taken soft. This gives rise to soft photon theorems which are universal at leading and sub-leading orders. We review it in appendix~\ref{softphoton}.
 
Now we want to derive the Ward identities associated to soft photons. We start with the following OPE between photon and matter fields which are minimally coupled:
 \be\label{photoope}
O_{\Delta,+1}(z_1,\bz_1) \cO_{\Delta',  J'} (z_2,\bz_2)
\sim \frac{e }{z_{12}} B(\Delta-1,\Delta'- J) O _{\Delta+\Delta'-1,  J'}(z_2,\bz_2)~.
\ee
This is very similar to the gluon case in \eqref{YMJ12} except that now we need to strip off the color index and use $O_{\Delta,+1}$. Also  $ O _{\Delta', J'} $ is a  matter  operator     with electric charge $e$ under  U(1) Maxwell field.

Repeating the same procedure, we obtain the Ward identities
 \beqn
&&
\EV{K^l(z , \bar z)   \cO_{\Delta_1,J_1} (z_1,\bar z_1)\cdots \cO_{\Delta_m,J_m} (z_m,\bar z_m)}
\nonumber\\&=& 
\sum_{k=1}^m e_k \sum_{s=0}^{1-l}   \frac{(\bz-\bz_k)^{s }}{z-z_k}   \;
 \frac{(-1)^{-s+1-l}}{s!(-s+1-l)!}    \frac{ \Gamma( 2\bar h_k)}{\Gamma( 2\bar h_k+l+s-1)}
 \nonumber \\&&
\times \bar\p_k^s  \EV{  \cO_{\Delta_1,J_1} (z_1,\bar z_1)\cdots \cO_{ \Delta_k+l-1,J_k} (z_k, \bar z_k )\cdots \cO_{\Delta_m,J_m} (z_m,\bar z_m)}~.
\label{softphotonWd}
 \eeqn
 which has similar structure as in the gluon case \eqref{gluonWds}. The   difference is that Maxwell theory itself is free and we need extra matter fields to interact. 
 
 For $l=1,0$, these Ward identities exactly coincide with the   leading and sub-leading soft photon theorem in \eqref{softphotontheorem} after performing Mellin transformation.

  We can also obtain the chiral Ward identities.  
  For leading soft photon, we find
    \beqn\label{ledingphotonder}
&&\EV{ \cK_0^1(z ) \cO_{\Delta_1,J_1}  (z_1,\bar z_1)\cdots \cO_{\Delta_m,J_m}  (z_m,\bar z_m)}
\\&=& 
\sum_{k=1}^m \frac{ e_k}{z-z_k  } 
     \EV{\cO_{\Delta_1,J_1} (z_1,\bar z_1)  \cdots O _{  \Delta_k-1, J_k} (z_k, \bar z_k )  \cdots \cO_{\Delta_m,J_m}  (z_1,\bar z_m)}~,
     \nonumber
\eeqn
which was also previously obtained in  \cite{Nande:2017dba} by different means. 

  For sub-leading soft photon,  we have two chiral Ward identities
  \beqn\label{subphotonder}
&&\EV{ \cK^{\frac32  }_{\frac12} (z ) \cO_{\Delta_1,J_1}  (z_1,\bar z_1)\cdots \cO_{\Delta_m,J_m}  (z_m,\bar z_m)}
\\&=& 
\sum_{k=1}^m \frac{- e_k}{z-z_k  }
 \Big[ (2\bar h_k-1)+\bz_k  \bar\p_k  \Big]
     \EV{\cO_{\Delta_1,J_1} (z_1,\bar z_1)  \cdots O _{  \Delta_k-1, J_k} (z_k, \bar z_k )  \cdots \cO_{\Delta_m,J_m}  (z_m,\bar z_m)}~,
     \nonumber
\eeqn
and
  \beqn\label{subphotonder2}
&&\EV{\cK^{\frac32  }_{-\frac12} (z ) \cO_{\Delta_1,J_1}  (z_1,\bar z_1)\cdots \cO_{\Delta_m,J_m}  (z_m,\bar z_m)}
\\&=& 
\sum_{k=1}^m \frac{ e_k}{z-z_k  }
 \bar\p_k   
     \EV{\cO_{\Delta_1,J_1} (z_1,\bar z_1)  \cdots O _{  \Delta_k-1, J_k} (z_k, \bar z_k )  \cdots \cO_{\Delta_m,J_m}  (z_m,\bar z_m)}~.
     \nonumber
\eeqn
 In \eqref{subphotonder2}, if we take a further derivative with respect to $z$ and thus consider the Ward identity associated with $\p\cK^{\frac32  }_{-\frac12} $, then we rediscover the Ward identity found in \cite{Himwich:2019dug}, which was shown to arise from  the sub-leading soft photon theorem.

 \paragraph{Magnetic corrections.} In the above discussions, we considered the matter particles which are electrically charged under U(1) Maxwell field. 
   It is known that there are magnetic corrections to the soft photon theorem as we review  in  appendix~\ref{softphoton}. Such corrections have also be understood from the perspective of asymptotic symmetry~\cite{Strominger:2015bla}. It is then natural to ask can we also include the magnetic corrections in our formalism? This turns out  to be very easy: we just need to complexify the couplings by replacing $e_k\to e_k+i g_k$ where $e_k,g_k$ are the electric and magnetic charges of particles.
  \footnote{~The replacing $e_k\to e_k+i g_k$ is for positive helicity photon; for negative helicity photon, we should replace $e_k\to e_k-i g_k$.  } 
    Indeed, in the three-point amplitude between photon and charged matter particle, nothing  prevents us from considering a complexified coupling for this three-point amplitude. A complexified coupling just means that the charged particle is a dyon with both electric   and magnetic charge. This can be easily understood from electromagnetic duality \footnote{~See  \cite{Nande:2017dba} for some discussions about electromagnetic duality  in this context.} which rotates the  phase of  the wave function of photon with definite helicity; as a result, the three-point coupling also acquires a phase. 
   This simple modification does not affect our derivation above at all except that we need to consider  complexified coupling  $e_k\to e_k+i g_k$ in OPE \eqref{photoope} and  \eqref{softphotonWd}. The resulting Ward identities with magnetic corrections  are thus similarly derived purely from 2D CCFT and are equivalent to the soft photon theorems. 
    
\section{Shadow Ward identities} \label{ShadowWardId}

In the last section, we derived the infinitely many Ward identities associated to the infinite dimensional soft currents. In this section, we would like to use shadow transformation to derive the  infinitely many shadow Ward identities. 

Generally, the shadow transformation of operator $O$ with weights $(h,\bar h)$  is defined as \cite{Osborn:2012vt} \footnote{ In this section and appendix \ref{integral}, for computational convenience we will treat $z, \bar z$ as complex conjugate of each other. }
\beqn\label{shadowntsf}
\widetilde O(w,\bar w) \equiv{\bf S}[O](w,\bar w)  &=&\int d^2 z\; (z-w)^{2h-2}(\bz-\bar w)^{2\bar h-2}O(z,\bar z)   ~.
\eeqn
As such, the shadow operator $\widetilde O$ has holomorphic and anti-holomorphic weights $(1-h,1-\bar h)$, or equivalently  conformal dimension  $2-\Delta$ and spin $-J$.

 Compared to Ward identities themselves, the shadow Ward identities play an equally important role in celestial holography. For example, by performing the   shadow transformation on the subleading soft graviton current, one gets the stress tensor in celestial CFT \cite{Cheung:2016iub,Kapec:2016jld}; and the shadow Ward identity at subleading order just coincides with the standard stress tensor  Ward identity.  We will generalize this construction and derive the shadow Ward identity associated with all the soft currents for graviton, gluon and photon.    Furthermore, a general formula of shadow Ward identities will be derived in  \eqref{generalWdId},  which is also applicable to soft gravitino and  soft gluino.  
%
%
%
%

\subsection{Shadow graviton Ward identity  }
The graviton Ward identities are derived in \eqref{gWd}. Let us write down again:
 \beqn
&&\EV{H^l(z , \bar z)   \cO_{\Delta_1,J_1} (z_1,\bar z_1)\cdots \cO_{\Delta_m,J_m} (z_m,\bar z_m)}
\nonumber \\&=&\sum_{k=1}^m \sum_{s=0}^{1-l}
\nonumber   \frac{(\bz-\bz_k)^{s+1}}{z-z_k} 
 \frac{(-1)^{-s -l}}{s!(-s+1-l)!}    \frac{ \Gamma(2\bar h_k+1)}{\Gamma(2\bar h_k+l+s )}
  \nonumber \\&&\qquad\times
     \bar\p_k^s \EV{  \cO_{\Delta_1,J_1} (z_1,\bar z_1)\cdots\cO_{ \Delta_k+l,  J_k}(z_2,\bar z_2 )\cdots \cO_{\Delta_m,J_m} (z_m,\bar z_m)}~.
     \label{gWd2}
 \eeqn
Note that soft graviton current $H^l$ has weight $(h,\bar h)=(\frac{l+2}{2},\frac{l-2}{2})$.  Its shadow current is  given by \eqref{shadowntsf}:
\be
\widetilde {H^l}(w,\bar w) =\int d^2 z\;  (z-w)^l (\bar z-\bar w)^{l-4} H^l(z,\bz)~.
\ee

To derive the shadow Ward identities, we need to perform a similar integral on the right hand side of original Ward identities \eqref{gWd2}. 

The integral has been    computed in   appendix~\ref{integral}. In particular, specializing  \eqref{IsABres} to the present case, we have
\be
\int d^2 z (z-w)^l (\bar z-\bar w)^{l-4}   \frac{(\bz-\bz')^{s+1 }}{z-z'} 
=\frac{(-1)^s  \pi (s+1)!}{(l-3) (l-2) \cdots (l+s-2 )} ( z'-w)^l (  \bz' -\bar w)^{l+s-2}~.
\ee

%
%

 Plugging this integral into \eqref{gWd2} and doing some algebra yields the following shadow Ward identities:
  \footnote{~It is interesting to note that the holomorphic dependence on shadow current is simply $( w- z_k )^l $. So for $l\le -1$, we may integrate $w$ and reduce    $( w- z_k )^l $ to $ 1/( w- z_k )$. This simplifies the equation a little bit but the physical meaning is not clear.
 }
   \beqn
&&\EV{\widetilde {H^l}(w, \bar w)   \cO_{\Delta_1,J_1} (z_1,\bar z_1)\cdots \cO_{\Delta_m,J_m} (z_m,\bar z_m)}
\nonumber \\&=&\frac{(-1)^{l }  \pi  }{(3-l)!   }   \sum_{k=1}^m \sum_{s=0}^{1-l}
 \frac{(s+1)  \Gamma( 2\bar h_k+1)}{\Gamma( 2\bar h_k+l+s )}
\times   ( w- z_k )^l ( \bar w- \bz_k )^{l+s-2}
 \nonumber    \\&&\qquad\times\;
     \bar\p_k^s \EV{  \cO_{\Delta_1,J_1} (z_1,\bar z_1)\cdots\cO_{ \Delta_k+l,  J_k}(z_k,\bar z_k )\cdots \cO_{\Delta_m,J_m} (z_m,\bar z_m)}~. 
 \eeqn
 
 Let us write down the identities at several leading orders explicitly.
 
For leading soft graviton  $l=1$, we have 
     \beqn
&&\EV{   \widetilde {H^1}(w, \bar w)   \cO_{\Delta_1,J_1} (z_1,\bar z_1)\cdots \cO_{\Delta_m,J_m} (z_m,\bar z_m)}
\nonumber \\&=& 
 -\frac\pi 2 \sum_{k=1}^m
\frac{w-z_k}{\bar w-\bz_k} 
  \EV{  \cO_{\Delta_1,J_1} (z_1,\bar z_1)\cdots \cO_{ \Delta_k+1,  J_k}(z_k,\bar z_k )\cdots \cO_{\Delta_m,J_m} (z_m,\bar z_m)}~.
 \eeqn
 This coincides with the Ward identity associated with  the  leading soft  (\emph{negative} helicity) graviton theorem.
 
For sub-leading soft graviton  $l=0$, we have 
    \beqn
&&\EV{\frac{3 }{  \pi  }   \widetilde{ H^0 } (   \bar w)   \cO_{\Delta_1,J_1} (z_1,\bar z_1)\cdots \cO_{\Delta_m,J_m} (z_m,\bar z_m)}
\nonumber \\&=&  \sum_{k=1}^m
 \Big[ \frac{\bar h_k}{(\bar w-\bar z_k)^2} +\frac{\bar\p_k}{\bar w-\bz_k}\Big] 
  \EV{  \cO_{\Delta_1,J_1} (z_1,\bar z_1)\cdots \cO_{\Delta_m,J_m} (z_m,\bar z_m)}~.
  \label{stresstensorWd}
 \eeqn
 This becomes the standard Ward identity of \emph{ anti-holomorphic } stress tensor in CFT   once we identify  $\bar T(\bar w)=\frac{3 }{  \pi  }   \widetilde {H^0}(   \bar w)  $. This stress tensor Ward identity  was previously discussed in  \cite{Cheung:2016iub,Kapec:2016jld, Fotopoulos:2019tpe}.%
 
For sub-sub-leading soft graviton  $l=-1$, we find 
    \beqn
&&\EV{    \widetilde{ H^{-1}} (   w,\bar w)   \cO_{\Delta_1,J_1} (z_1,\bar z_1)\cdots \cO_{\Delta_m,J_m} (z_m,\bar z_m)}
=
 -\frac{\pi}{4!}
  \sum_{k=1}^m \frac{1}{w-z_k}
   \\& & \qquad \times
 \Big[ \frac{2\bar h_k(2\bar h_k-1)}{(\bar w-\bar z_k)^3} +\frac{4\bar h_k\bar\p_k}{(\bar w-\bar z_k)^2} 
 +\frac{3\bar\p_k^2}{\bar w-\bz_k}\Big] 
  \EV{  \cO_{\Delta_1,J_1} (z_1,\bar z_1)\cdots \cO_{ \Delta_k-1,  J_k}(z_k,\bar z_k )\cdots \cO_{\Delta_m,J_m} (z_m,\bar z_m)}~.
  \nonumber
 \eeqn

\subsection{Shadow photon and gluon Ward identity  }

Now we switch to shadow Ward identities  for soft photons. The  soft photon Ward identities are given  in \eqref{softphotonWd}:
 \beqn
&&
\EV{K^l(z , \bar z)   \cO_{\Delta_1,J_1} (z_1,\bar z_1)\cdots \cO_{\Delta_m,J_m} (z_m,\bar z_m)}
\nonumber\\&=& 
\sum_{k=1}^m e_k \sum_{s=0}^{1-l}   \frac{(\bz-\bz_k)^{s }}{z-z_k}   \;
 \frac{(-1)^{-s+1-l}}{s!(-s+1-l)!}    \frac{ \Gamma( 2\bar h_k)}{\Gamma( 2\bar h_k+l+s-1)}
 \nonumber \\&&
\times \bar\p_k^s  \EV{  \cO_{\Delta_1,J_1} (z_1,\bar z_1)\cdots \cO_{ \Delta_k+l-1, J_k} (z_k, \bar z_k )\cdots \cO_{\Delta_m,J_m} (z_m,\bar z_m)}~.
\label{softphotonWd2}
 \eeqn

The soft photon current $K^l$ has weight $(h,\bar h)=(\frac{l+1}{2},\frac{l-1}{2})$. Thus the soft     photon  shadow current  is given by  \eqref{shadowntsf}:
\be
\widetilde {K^l} (w,\bar w) =\int d^2 z\; (z-w)^{l-1} (\bar z-\bar w)^{l-3} {K^l} (z,\bz)~.
\ee

To proceed, we   need to use the following integral which has been derived  in \eqref{IsABres}:
\be
\int d^2 z (z-w)^{l-1} (\bar z-\bar w)^{l-3} \frac{(\bz-\bz')^{s  }}{z-z'} 
=\frac{(-1)^s (- \pi) s!}{(l-2) (l-2) \cdots (l+s-2 )} ( z'-w)^ {l-1} (  \bz' -\bar w)^{l+s-2}~. \quad
\ee
 
%
%

 Plugging this integral into \eqref{softphotonWd2}  gives the  following shadow Ward identities:
\beqn
&&
\EV{\widetilde {K^l}(w,\bar w)   \cO_{\Delta_1,J_1} (z_1,\bar z_1)\cdots \cO_{\Delta_m,J_m} (z_m,\bar z_m)}
\nonumber \\&=&  \frac{(-1)^{ -l}\pi}{(2-l)!} 
\sum_{k=1}^m e_k \sum_{s=0}^{1-l} 
    \frac{ \Gamma( 2\bar h_k)}{\Gamma( 2\bar h_k+l+s-1)}
  ( w-z_k )^ {l-1} (  \bar w-\bz_k )^{l+s-2}
\nonumber \\&&\qquad\qquad
\times  
\bar\p^s  \EV{  \cO_{\Delta_1,J_1} (z_1,\bar z_1)\cdots \cO_{ \Delta_k+l-1,J_k} (z_k, \bar z_k )\cdots \cO_{\Delta_m,J_m} (z_m,\bar z_m)}~.
 \eeqn
 
 More explicitly for leading soft photon $l=1$, we have 
\beqn
&&
\EV{\widetilde {K^1}(w,\bar w)   \cO_{\Delta_1,J_1} (z_1,\bar z_1)\cdots \cO_{\Delta_m,J_m} (z_m,\bar z_m)}
\nonumber\\&=& -\pi
\sum_{k=1}^m  
\frac{e_k }{\bar w-\bz_k}  \EV{  \cO_{\Delta_1,J_1} (z_1,\bar z_1)\cdots    \cO_{\Delta_m,J_m} (z_m,\bar z_m)}~,
 \eeqn
which is the same as the Ward identity associated with  the  leading soft   photon with negative helicity.

 While for sub-leading soft photon   $l=0$, we have 
\beqn\label{subphoton}
&&
\EV{\widetilde {K^0}(w,\bar w)   \cO_{\Delta_1,J_1} (z_1,\bar z_1)\cdots \cO_{\Delta_m,J_m} (z_m,\bar z_m)}
\\&=& \frac{\pi}{2}
\sum_{k=1}^m  
\frac{e_k }{  w- z_k}\Big[\frac{2\bar h_k -1}{(\bar w-\bz_k)^2} +\frac{\bar\p_k}{\bar w-\bz_k}  \Big]   \EV{  \cO_{\Delta_1,J_1} (z_1,\bar z_1)\cdots   \cO_{ \Delta_k -1,J_k} (z_k, \bar z_k )\cdots   \cO_{\Delta_m,J_m} (z_m,\bar z_m)}~. 
\nonumber
 \eeqn

 
 For gluon,  their corresponding shadow Ward identities are similarly given by:
 \beqn
&&
\EV{\widetilde {K^{l,a}}(w,\bar w)   \cO^{b_1}_{\Delta_1,J_1} (z_1,\bar z_1)\cdots \cO^{b_m}_{\Delta_m,J_m} (z_m,\bar z_m)}
\nonumber\\&=&  \frac{(-1)^{ -l}\pi}{(2-l)!} 
\sum_{k=1}^m f^{ab_k c_k}  \sum_{s=0}^{1-l} 
    \frac{ \Gamma( 2\bar h_k)}{\Gamma( 2\bar h_k+l+s-1)}
  ( w-z_k )^ {l-1} (  \bar w-\bz_k )^{l+s-2}
\nonumber \\&&\qquad\qquad
\times  
\bar\p^s  \EV{  \cO^{b_1}_{\Delta_1,J_1} (z_1,\bar z_1)\cdots \cO^{c_k}_{ \Delta_k+l-1,J_k} (z_k, \bar z_k )\cdots \cO^{b_m}_{\Delta_m,J_m} (z_m,\bar z_m)}~.
 \eeqn

\section{EFT corrections} \label{EFT}

In the previous sections, we proposed a method for deriving Ward identities from celestial OPE.  With this procedure, we discussed the  Ward identities  associated with the soft symmetry currents in supersymmetric EYM theory. Although we were considering this specific theory, the Ward identities 
are supposed to  hold more generally as they capture the universal feature of quantum fields. However, they are not always universal; there are various types of correction  to the soft theorems due to quantum loops or higher derivative interactions. More specifically, it has been shown in \cite{Elvang:2016qvq} that there are cubic vertices which can modify the sub-sub-leading soft graviton theorem and sub-leading soft photon theorem. 
These claims are derived in \cite{Elvang:2016qvq} by considering local unitary effective field theory and analyzing all possible local operators. Therefore,    the leading and sub-leading soft graviton theorems, as well as the leading soft photon theorem  are indeed universal   at tree level  in EFT as  guaranteed by locality and unitarity. \footnote{~The Weinberg's leading soft graviton theorem is even robust against quantum loops, while the rest of soft theorems may suffer from quantum corrections.  }

We will reformulate their results in the language of celestial holography.  More specifically, we will consider the EFT corrections to our previous Ward identities. The procedure is the same as that in the previous sections. We will first derive a general celestial OPE \eqref{opeeft} arising from the cubic interaction of three massless spinning particles.  Based on this general OPE, we establish the general Ward identities \eqref{generalWdId} and its shadow cousin \eqref{generalSdWdId}.
Applying the general results to EFT, we find that the  corrections to Ward identities indeed  start to appear at sub-sub-leading order for soft graviton and sub-leading order for soft photon. 

\subsection{General celestial OPE  and Ward identity } \label{genope11}
As we derived in appendix~\ref{opeEFT}, the  leading tree level celestial OPEs   arising from cubic    vertices  of three spinning massless particles take the following general form: 
 \be\label{opeeft}
\cO_{\Delta_1, J_1}(z_1, \bz_1)\cO_{\Delta_2, J_2}(z_2, \bz_2)
\sim  \kappa_{J_1J_2 J_3} \frac{ \bz_{12}^{J_1+J_2+J_3-1}}{z_{12} }
B\Big(\Delta_1+J_2+J_3-1,\Delta_2+J_1+J_3-1 \Big) 
\cO_{\Delta_3, -J_3}(z_2, \bz_2)~, \qquad
\ee
where $\Delta_3=\Delta_1+\Delta_2+J_1+J_2+J_3-2$ and $\kappa_{J_1J_2 J_3}$ is the coupling constant of the cubic vertex. This is derived in condition $\cJ\equiv J_1+J_2+J_3\ge 0$. \footnote{~In case $\cJ\le 0$, a similar OPE can be obtained by flipping the spin $J_i\to -J_i$ and exchanging $z_i\leftrightarrow \bar z_i$.}
One can check that this OPE agrees with all known celestial OPEs, including those in \eqref{ope1}-\eqref{ope10}.
 
 As before, we also want to sum  over  all the $\overline{SL(2,\mathbb R)}$  descendants. Using \eqref{OPEdes}, we get 
  \beqn
&&  \cO_{\Delta_1, J_1}(z_1, \bz_1)\cO_{\Delta_2, J_2}(z_2, \bz_2)
\nonumber \\&\sim&
 \kappa_{J_1J_2 J_3} \frac{ \bz_{12}^{J_1+J_2+J_3-1}}{z_{12} }
\int_0^1 dt\; \cO_{\Delta_3, -J_3}(z_2, \bz_2+t\bz_{12})\;
t^{\Delta_1+J_2+J_3-2}(1-t)^{\Delta_2+J_1+J_3-2}
~.\qquad
\eeqn
Doing the integral  thus gives the OPE where all the $\overline{SL(2,\mathbb R)}$  descendant contributions are included: 
  \beqn
&&  \cO_{\Delta_1, J_1}(z_1, \bz_1)\cO_{\Delta_2, J_2}(z_2, \bz_2)
\nonumber\\&\sim&
  \frac{ \kappa_{J_1J_2 J_3}}{z_{12} }
  \sum_{s=0}^\infty\frac{(\bz_{12})^{\cJ+s-1}}{s!} \bar\p^s \cO_{\Delta_3, -J_3}(z_2, \bz_2 )\;
B(\Delta_1+s+J_2+J_3-1,\Delta_2+J_1+J_3-1)
~. \qquad
\eeqn

Setting $\Delta_1\to k$ where $k\in \mathbb Z $ for bosonic soft current or  $k\in \mathbb Z+\frac12 $ for fermionic soft current, we obtain the OPE between soft currents and hard operators 
  \beqn\label{softhardgenearlOPE}
&&  R^{k, J_1}(z_1, \bz_1) \cO_{\Delta_2, J_2}(z_2, \bz_2)
\\&\sim&
  \frac{\kappa_{J_1J_2 J_3}}{z_{12} }
  \sum_{s=0}^ {1-k-J_2-J_3} \frac{(\bz_{12})^{\cJ+s-1}}{s!} \bar\p^s \cO_{\Delta_3, -J_3}(z_2, \bz_2 )\;
\frac{(-1)^{(1-k-s-J_2-J_3)}}{ (1-k-s-J_2-J_3)! }
\frac{ \Gamma(  \Delta_2+J_1+J_3-1) }{\Gamma( \Delta_2+J_1+J_2+2J_3+k+s -2)}
~, \nonumber
\eeqn
where we used \eqref{Betapole2}. The Ward identity can then be easily established by replacing each pair of soft current and hard operator with their OPEs above. 

Explicitly, the general formula of Ward identities is given by 
  \beqn\label{generalWdId}  
&& \EV{ R^{l, J }(z,\bz) \cO_{\Delta_1, J_1}(z_1, \bz_1)\cdots \cO_{\Delta_m, J_m}(z_m, \bz_m) }
\nonumber \\&=&
 \sum_{k=1}^m
 \kappa_{JJ_kJ_k'}  (-1)^{\nu(\nu_1+\cdots +\nu_{k-1})} 
  \sum_{s=0}^ {1-l-J_k-J_k'}  
\frac{(\bz-\bz_k)^{J+J_k+J_k'+s-1}}{ z -z_k } 
\nonumber \\&&\qquad\qquad\qquad
\times
  \frac{(-1)^{ (1-l-s-J_k-J_k')}}{ s! (1-l-s-J_k-J_k')! }
\frac{ \Gamma(  \Delta_k+J +J_k'-1) }{\Gamma( \Delta_k+J +J_k+2J_k'+l+s -2)}
\nonumber \\&&\qquad\qquad\qquad \times
\bar\p_k^s \EV{   \cO_{\Delta_1, J_1}(z_1, \bz_1)\cdots \cO_{\Delta_k+l+J+J_k+J_k' -2 , -J_k'}(z_k, \bz_k )\cdots  \cO_{\Delta_m, J_m}(z_m, \bz_m) 
}
~,\qquad
\eeqn
and its corresponding shadow cousin  is
  \beqn\label{generalSdWdId}
&& \EV{ \widetilde{R^{l, J }}(w,\bar w) \cO_{\Delta_1, J_1}(z_1, \bz_1)\cdots \cO_{\Delta_m, J_m}(z_m, \bz_m) }
\nonumber \\&=&
\frac{\pi (-1)^{J+l+1}}{(1+J-l)!}
 \sum_{k=1}^m    \kappa_{JJ_kJ_k'}  (-1)^{\nu(\nu_1+\cdots +\nu_{k-1})}  \sum_{s=0}^ {1-l-J_k-J_k'} 
 (w-z_k)^{l+J-2} (\bar w-\bar z_k)^{J_k+J_k'+l+s-2}
\nonumber \\&&  \qquad\qquad\qquad  \times\;
\frac{(J+J_k+J_k'+s-1) !}{s!}
\frac{  \Gamma(  \Delta_k+J +J_k'-1) }{\Gamma( \Delta_k+J +J_k+2J_k'+l+s -2)}
\nonumber \\&&  \qquad\qquad\qquad
\times\;
  \bar\p_k^s \; \EV{   \cO_{\Delta_1, J_1}(z_1, \bz_1)\cdots \cO_{\Delta_k+l+J+J_k+J_k' -2 , -J_k'}(z_k, \bz_k )\cdots  \cO_{\Delta_m, J_m}(z_m, \bz_m) 
}
~.\qquad
\eeqn
where to take into account the statistics we introduce $\nu_i=(2 J_i) \mod 2=0, 1$ for bosonic and fermionic operators, respectively.  The chiral Ward identities can also be easily obtained by performing mode expansion using \eqref{Rk} and \eqref{modeRedefine}.

This establishes the general celestial OPEs and Ward identities. 
Our discussions in the previous sections just correspond  to the special case of the above formulae. 
For soft graviton, we have $J=2, J_k=-J_k'$; for soft gluon/photon, we have $J=1, J_k=-J_k'$.
These formulae are also applicable to fermionic soft current. In the case of minimal coupling, we have 
$J=3/2, J_k-1/2=-J_k'$ for soft gravitino, and  $J=1/2, J_k-1/2=-J_k'$ for soft gluino/photino. Inserting these values to the above formulae, we get infinitely many   Ward identities corresponding to the fermionic symmetries. 

 In the next two subsections, we will use the general results here to discuss the corrections to Ward identities for sub-leading soft photon and sub-sub-leading soft graviton based on~\cite{Elvang:2016qvq}. 
%
%

\subsection{EFT correction to photon Ward identity}

Let us first discuss the case of photon   $J_1=1$. Following \cite{Elvang:2016qvq}, the EFT corrections   appear  at $J_2+J_3=1$. The corresponding celestial OPE reads \eqref{opeeft}:
 \be
 \cO_{\Delta_1, +1}(z_1,\bz_1)  \cO_{\Delta_2,  J_2} (z_2,\bz_2)\sim  \frac{\bz_{12} }{z_{12} }  B(\Delta_1, \Delta_2-J_2+1) 
 \cO_{\Delta_1+\Delta_2,  J_2-1}~,
 \ee
 where we suppress the coupling constant  for simplicity. 
  Specializing \eqref{softhardgenearlOPE} to the present case, we get the OPE between soft photon current and hard operator   
   \beqn
K^k(z_1,\bz_1)  \cO_{\Delta_2,  J_2} (z_2,\bz_2)
  &\sim&  \frac{\bz_{12} }{z_{12} } 
 \sum_{s=0}^ {-k}\frac{(\bz_{12})^s}{s!}\bar\p^sO_{ \Delta_2+k,  J_2-1}(z_2,\bz_2  )
\frac{(-1)^{-k-s}}{(-k-s)!} \frac{\Gamma(\Delta_2-J_2+1)}{\Gamma(\Delta_2-J_2+k+s+1)}~.
  \qquad\qquad
 \eeqn
 
 The resulting Ward identity is
          \beqn\label{photoncorrection}
&&\EV{K^l (z,\bar z  )  \cO_{\Delta_1,  J_1} (z_1,\bz_1) \cdots \cO_{\Delta_m,  J_m} (z_m,\bz_m)}
\nonumber\\&=&
     \sum_{k=1}^m 
     \sum_{s=0}^ {-l} \frac{(\bz-\bz_k)^{s+1}}{z-z_k }  
\frac{(-1)^{-l-s}}{s!(-l-s)!} \frac{\Gamma(\Delta_k-J_k+1)}{\Gamma(\Delta_k-J_k+l+s+1)} 
\nonumber\\& &  \qquad\qquad
\times  \bar\p_k^{s}
\EV{\cO_{\Delta_1,  J_1} (z_1,\bz_1) \cdots  \cO_{\Delta_k+l,  J_k-1}(z_k,\bz_k  )
\cdots \cO_{\Delta_m,  J_m} (z_m,\bz_m)}~. \quad
   \eeqn
  Note that in practice we need to combine the Ward identities for     all  the interactions together. 
    
   Let us look at \eqref{photoncorrection} in more detail. At leading order $l=1$, we see it has no effect.  At sub-leading order $l=0$, it become non-trivial  
        \beqn
&&\EV{ K^{0} (z,\bz)  \cO_{\Delta_1,  J_1} (z_1,\bz_1) \cdots \cO_{\Delta_m,  J_m} (z_m,\bz_m)}
\nonumber\\&=&
 \sum_{k=1}^m\frac{\bz-\bz_k}{z-z_k }  
\EV{\cO_{\Delta_1,  J_1} (z_1,\bz_1) \cdots  \cO_{\Delta_k ,  J_k-1}(z_k,\bz_k  )
\cdots \cO_{\Delta_m,  J_m} (z_m,\bz_m)}~,
   \eeqn
   which  indeed  agree with corrections to soft theorem found in \cite{Elvang:2016qvq} after a Mellin transformation.

   In terms of chiral soft photon currents, the corresponding chiral Ward identities are:
         \beqn
&&\EV{\cK_n^i (z  )  \cO_{\Delta_1,  J_1} (z_1,\bz_1) \cdots \cO_{\Delta_m,  J_m} (z_m,\bz_m)}
\nonumber\\&=&
     \sum_{k=1}^m\frac{1}{z-z_k }  
   \sum_{r=\max(0,n+2-i)}^ { n+i-1}   \PBK{i+n-1\\r}   
   \frac{   {(-1)^{ i +n-1 }}   (r+i-n-1)  \Gamma(\Delta_k-J_k+1)}{\Gamma(\Delta_k-J_k +r-i-n+2)}
\nonumber\\& &  \qquad\qquad
\times \bar z_k ^r \bar\p_k^{r+i-n-2}
\EV{\cO_{\Delta_1, J_1} (z_1,\bz_1) \cdots  \cO_{\Delta_k-2i+3,  J_k-1}(z_k,\bz_k  )
\cdots \cO_{\Delta_m,  J_m} (z_m,\bz_m)}~. \qquad
   \eeqn
  
\subsection{EFT correction to graviton Ward identity}

  Now we switch to the case of graviton $J_1=2$. Following \cite{Elvang:2016qvq}, the EFT corrections   appear  at  $J_2+J_3=2$. The corresponding celestial OPE reads \eqref{opeeft}:
 \be
 \cO_{\Delta_1, +2} (z_1 ,\bz_1) \cO_{\Delta_2,  J_2} (z_2,\bz_2)\sim  \frac{\bz_{12} ^3}{z_{12} }  B(\Delta_1+1, \Delta_2-J_2+3) 
 \cO_{\Delta_1+\Delta_2+2,  J_2-2}(z_2,\bz_2) ~,
 \ee 
where the EFT coupling is again suppressed. 

Now the soft graviton current has  OPE \eqref{softhardgenearlOPE} 
   \beqn
H^k(z_1,\bz_1)  \cO_{\Delta_2,  J_2} (z_2,\bz_2)
 &\sim&  
  \frac{\bz_{12}^3 }{z_{12} } 
  \sum_{s=0}^ {-k-1}\frac{(\bz_{12})^s}{s!}\bar\p^sO_{ \Delta_2+k+2,  J_2-2}(z_2,\bz_2 ) 
  \frac{(-1)^{-k-s-1}  }{(-k-s-1)!}  \frac{  \Gamma(\Delta_2-J_2+3)}{\Gamma(\Delta_2-J_2+k+s+4)}~.
\nonumber\\  
 \eeqn
  The resulting Ward identity is
          \beqn\label{gravitoncorreection}
&&\EV{H^l (z,\bar z  )  \cO_{\Delta_1,  J_1} (z_1,\bz_1) \cdots \cO_{\Delta_m,  J_m} (z_m,\bz_m)}
\nonumber\\&=&
     \sum_{k=1}^m 
     \sum_{s=0}^ {-l-1} \frac{(\bz-\bz_k)^{s+3}}{z-z_k }  
\frac{(-1)^{-l-s-1}}{s!(-l-s-1)!} \frac{\Gamma(\Delta_k-J_k+3)}{\Gamma(\Delta_k-J_k+l+s+4)} 
\nonumber\\& &  \qquad\qquad
\times  \bar\p_k^{s}
\EV{\cO_{\Delta_1,  J_1} (z_1,\bz_1) \cdots  \cO_{\Delta_k+l+2,  J_k-2}(z_k,\bz_k  )
\cdots \cO_{\Delta_m,  J_m} (z_m,\bz_m)}~.
   \eeqn
   
At leading and sub-leading order $l=1,0$, it has no effect.  At sub-sub-leading order $l=-1 $, we find 
            \beqn
&&\EV{    H^{-1}  (z ,\bz )    \cO_{\Delta_1,  J_1} (z_1,\bz_1) \cdots \cO_{\Delta_m,  J_m} (z_m,\bz_m)}
\nonumber\\&=&
     \sum_{k=1}^m\frac{  (\bar z-\bz_k) ^3}{z-z_k }  
\EV{\cO_{\Delta_1,  J_1} (z_1,\bz_1) \cdots  \cO_{\Delta_k+1,  J_k-2}(z_k,\bz_k  )
\cdots \cO_{\Delta_m,  J_m} (z_m,\bz_m)}~.
   \eeqn
One can show that this is consistent with the corrections to soft graviton theorem found in \cite{Elvang:2016qvq} after doing a Mellin transformation.
%

   As before we can also establish chiral Ward identities: 
         \beqn
&&\EV{\cH_n^i (z  )  \cO_{\Delta_1,  J_1} (z_1,\bz_1) \cdots \cO_{\Delta_m,  J_m} (z_m,\bz_m)}
\nonumber\\&=&
     \sum_{k=1}^m\frac{1}{z-z_k }  
   \sum_{r=\max(0,n+4-i)}^ { n+i-1}      {(-1)^{ i +n+1 }}    \PBK{i+n-1\\r} 
  \frac{(r+i-n-1)(r+i-n-2)(r+i-n-3 )\Gamma(\Delta_2-J_2+3)}{\Gamma(\Delta_2-J_2 +r-i-n+4)}
\nonumber\\& &  \qquad\qquad
\times  \bar z_k ^r \bar\p_k^{r+i-n-4}
\EV{\cO_{\Delta_1,  J_1} (z_1,\bz_1) \cdots  \cO_{\Delta_k-2i+6,  J_k-2}(z_k,\bz_k  )
\cdots \cO_{\Delta_m,  J_m} (z_m,\bz_m)}~.
   \eeqn
   
 \subsection{Absence of EFT correction to holographic chiral algebra}

Finally, we want to understand whether there are corrections to holographic chiral algebra based on the EFT framework and especially the result in   \cite{Elvang:2016qvq}.

In   \cite{Elvang:2016qvq}, the authors considered cubic vertices involving three massless particles,  and take one of them soft, which we will label   as  particle 1. Then they define $\beta=J_1-J_2-J_3+1$. Note the  mass dimension  of the coupling constant for this vertex is given by  $\beta-2J_1=1-\cJ$. Obviously, we have  $\beta, \cJ\in \mathbb Z$ in order to have a bosonic effective operator.

In \cite{Elvang:2016qvq}  it was argued that in local unitary EFT, any cubic vertex should satisfy  $\beta<4$. What's more, if the vertex involves   photons, a stronger condition is $\beta\le 2$.
  This immediately implies that no local cubic interactions involving photons, gravitinos, or gravitons are allowed if the sum of   their helicities vanishes. This means that in  our OPE \eqref{opeeft}, the pole of the form $1/(z_{12}\bz_{12})$ is forbidden in gravitational or Maxwell-matter theory. \footnote{~It would be very interesting to translate the principles of locality, unitarity and causality into some principles in celestial CFT. Then we may use the principles  in 2D   to establish these claims.}

When the soft  particle 1 is graviton, namely $J_1=2$, the condition $\beta<4$  gives $\beta=3,2,1,\cdots$. For $\beta=3$, we have $J_2+J_3=0$. This just corresponds to the universal gravitational coupling between graviton and matter fields,  whose celestial OPE is given in \eqref{gravityOPE}. 
 For $\beta=2$ and thus $J_2+J_3=1$, it was argued in \cite{Elvang:2016qvq}   that no effective operator satisfying  this condition is allowed. 
 For $\beta=1$ and thus $J_2+J_3=2$, the full list of   EFT operators is given by \cite{Elvang:2016qvq} 
   \be 
\phi R_{\mu\nu\rho\sigma}R^{\mu\nu\rho\sigma}~, \quad
R_{\mu\nu\rho\sigma} F^{\mu\nu}F^{\rho\sigma}~,\quad
R^{\mu\nu\rho\sigma}\bar\psi_\rho\gamma_{\mu\nu}\p_\sigma\chi~.
 \ee
 They can modify the sub-sub-leading soft graviton theorem and the Ward identities as we discussed before. 
  
When the soft  particle 1 is photon, namely $J_1=1$, the allowed values for $\beta$ are $\beta=2,1,\cdots$.  For $\beta=2$ and thus $J_2+J_3=0$, this is just  the minimal coupling between photon and charged matter, whose celestial OPE is described in \eqref{photoope}. For $\beta=1$ and thus $J_2+J_3=1$, the full list of allowed  effective field theory operators  is  \cite{Elvang:2016qvq} 
  \be\label{eftopphoton}
 \bar \chi \gamma^{\mu\nu}F_{\mu\nu}\chi~, \quad
 \phi F_{\mu\nu} F^{\mu\nu}~, \qquad
 \phi F_{\mu\nu} \tilde F^{\mu\nu}~, \qquad
  \bar\psi_\mu F_{\nu\rho} \gamma^{\mu\nu\rho}\chi~, \qquad
  h_{\mu\nu}\Big(F^{\mu\rho}F^{\nu} {}_\rho   -\frac14\eta^{\mu\nu} F_{\rho\sigma}F^{\rho\sigma}  \Big) ~.
 \ee
  This  type of vertex can modify the sub-leading soft photon theorem and Ward identity as we discussed already. 
 
 Now we want to see whether there are corrections to the holographic chiral algebra from the effective  field theory operators we listed above. Since our formalism can only deal with positive helicity soft particles, we need to restrict to  $J_1, J_2,-J_3>0$ in \eqref{opeeft}. It is easy to see that this is possible only for the last two effective operators in \eqref{eftopphoton}, which  have $J_2=3/2,J_3=-1/2$ and $J_2=2,J_3=-1$, respectively.  But these are just the gravitational coupling between gravity and photon, which we have considered already. Therefore, all the EFT operators up to this order have no effects on holographic chiral algebra. 
 
 One might still worry about even higher derivative operators at smaller $\beta$ which we did not consider above; they  may modify the Ward identities at even less leading orders.
 However, the holographic chiral algebra is generated by the soft graviton currents up to sub-sub-leading order and   soft photon currents up to sub-leading order. \footnote{~This comes from the fact that  celestial OPEs are fully determined by   soft theorems up to these orders  \cite{Pate:2019lpp}.} 
 Now that there are no EFT corrections at these several leading orders, the higher orders   should  also not be affected.  To conclude, in case we can apply our formalism, any effective operators can not modify the holographic chiral algebra.

 \section{Conclusion}\label{conclusion}
 
 To summarize, in this paper, we studied many aspects of symmetry in celestial holography by deriving the holographic chiral algebra and the  associated Ward identities. These symmetries are hidden in the traditional framework of S-matrix, but play important roles in   governing the consistency of quantum fields.  Inspired by \cite{Guevara:2021abz}, we established a   general  and systematic framework to reveal these hidden symmetries by making full use of the techniques in  conformal field theory. The input of this formalism is the celestial OPEs, while the output is an  infinite-dimensional chiral symmetry algebra as well as their corresponding Ward identities. Moreover, we also derived a general formula for   tree-level leading order celestial OPE arising from cubic interactions of three spinning massless particles. As a result, we found a general formula of Ward identities \eqref{generalWdId} and \eqref{generalSdWdId}.
 
 
 In spite of various results in this paper,  many   interesting questions   remain   to be further explored. Maybe the most important question is to understand the role of  soft  particles with negative helicity.
Solving  this question may finally enable us to discover the full symmetry algebra.
   
It would be  also  useful to understand the infinite Ward identities from the traditional  momentum space perspective. 
In \cite{Hamada:2018vrw,Li:2018gnc}, an infinite set of soft theorems was found   for photon and graviton. And in the  MHV sector of gravity, \cite{Guevara:2019ypd} also found infinitely many conformally soft theorems.  These soft theorems are likely to be equivalent to our Ward identities, but  it remains to check the equivalence explicitly.  
 
 Furthermore, the Ward identities we established in this paper have a single current insertion  and correspond  to the single soft theorems. However, there are also many types of double  and  multiple soft theorems in gauge theory and gravity. It is natural to ask whether we can also  establish the   Ward identities with multiple  current insertions which are equivalent to the multiple soft theorems. 
 
Another interesting direction is to generalize our framework here to  higher dimensions. Although the   soft theorems and collinear factorizations of amplitudes are  also  well studied in higher dimensions,  the  celestial holography in higher dimensions   is largely unexplored. Many results in 4D should be generalizable to higher dimensions.  In particular, the equivalence between sub-leading soft graviton theorem  and stress tensor Ward identity has been  established in~\cite{Kapec:2017gsg}.
This may be a good starting point for a systematic exploration of symmetry and Ward identity in higher dimensions.

Finally, it is vital to  fully understand the corrections to the holographic chiral algebra and Ward identities from quantum loops   and higher derivative interactions.   We attempted to address this question in EFT,
but  our analysis is restricted to  the  tree level case with  positive soft particles only. We still need to understand the negative soft particles. On the other hand, although   the sub-leading soft graviton theorem is not corrected in EFT, it nevertheless suffers quantum correction which has been shown to be one-loop exact \cite{Bern:2014oka}. \footnote{More precisely, the IR divergent part is proved  to be one-loop exact, while the IR finite part has only been shown to  be one-loop exact in some explicit examples. We thank Congkao Wen for discussion on this point. }  In \cite{He:2017fsb}, this one-loop exact soft theorem was further translated into the loop-corrected stress tensor Ward identity. Since in this paper we have established a systematic framework for Ward identities, it is thus very interesting to incorporate quantum corrections into our formalism and  then re-derive the loop-corrected stress tensor  Ward identity. This would also allow us to derive the deformation of the holographic chiral algebra.  In particular, in the graviton case, the classical $w_{1+\infty}$ algebra is supposed to be deformed to  some type  of $W_{1+\infty}$  algebra at quantum level.


 \acknowledgments
 We would like to thank  Ana-Maria Raclariu, Andrew  Strominger, and Tomasz   Taylor for correspondence.
This work was  supported by the Royal Society grant, \textit{“Relations, Transformations, and Emergence in Quantum Field Theory”}, and 
by the Science and Technology Facilities Council (STFC) Consolidated Grant ST/T000686/1 \textit{``Amplitudes, strings  \& duality''}.

\appendix

\section{Celestial super-OPEs in terms of celestial superfields}\label{superOPEs}
In \eqref{ope1}-\eqref{ope10}, we wrote down explicitly the celestial OPEs for all the component fields in supersymmetric EYM theory.  In this appendix  we would like to rewrite all the OPEs in a manifest supersymmetric way  by introducing the following celestial on-shell superfields \cite{Fotopoulos:2020bqj,Jiang:2021xzy}:
\beqn\label{superfd1}
U_\Delta^a(z,\bar z,\eta) &=&\cO^a_{\Delta,+1}(z,\bar z)+\eta \cO^a_{\Delta,+\frac12}(z,\bar z)~,
\\
W_\Delta(z,\bar z,\eta) &=&\cO _{\Delta,+2}(z,\bar z)+\eta \cO _{\Delta,+\frac32}(z,\bar z)~,
\label{superfd2}
\eeqn
where $\eta$ is the anti-commuting Grassmann variable. 

As in the case of  $\mathcal N=4$ SYM theory \cite{Jiang:2021xzy}, it is natural to write down the following super-OPEs:
\beqn
U_{\Delta_1}^a(z_1,\bar z_1,\eta_1) U_{\Delta_2}^b(z_2,\bar z_2, \eta_2) &\sim&
 \frac{f^{abc}}{z_{12}} U_{\Delta_1+\Delta_2-1}^c(z_2,\bar z_2, \eta_1 e^{\frac12\p_{\Delta_1}}+ \eta_2 e^{\frac12\p_{\Delta_2}}  )  B(\Delta_1-1,\Delta_2-1)~, 
\qquad\quad
\\
W_{\Delta_1} (z_1,\bar z_1,\eta_1) W_{\Delta_2} (z_2,\bar z_2, \eta_2) &\sim&
- \frac{\bar z_{12}}{z_{12}} W_{\Delta_1+\Delta_2 } (z_2,\bar z_2, \eta_1 e^{\frac12\p_{\Delta_1}}+ \eta_2 e^{\frac12\p_{\Delta_2}}  )  B(\Delta_1-1,\Delta_2-1)~, 
\qquad\quad
\\
U^a_{\Delta_1} (z_1,\bar z_1,\eta_1)  W_{\Delta_2} (z_2,\bar z_2, \eta_2) &\sim&
- \frac{\bar z_{12}}{z_{12}} U^a_{\Delta_1+\Delta_2 } (z_2,\bar z_2,\eta_1 e^{\frac12\p_{\Delta_1}}+ \eta_2 e^{\frac12\p_{\Delta_2}}  )  B(\Delta_1 ,\Delta_2-1)~. 
\qquad\quad
\eeqn
By expanding the Grassmann parameter, it is easy to verify that that the three super-OPEs above agree exactly with the component OPE  in \eqref{ope1}-\eqref{ope10}.

The  supersymmetry generators acting on these  celestial on-shell superfields are given by \cite{Fotopoulos:2020bqj,Jiang:2021xzy}:
\beqn\label{Qtsf}
Q^\alpha=  \PBK{1 \\ z}^\alpha e^{\frac12\p_\Delta} \eta~, \qquad
\tilde Q^{\dot\alpha}=  \PBK{1 \\  \bar z}^{\dot\alpha} e^{ \frac12\p_\Delta} \p_\eta~,
 \eeqn
which satisfy the standard commutation relation of supersymmetry algebra
 \be
 \{Q^\alpha, \; \tilde Q^{\dot\alpha} \}= P^{\alpha \dot\alpha}~, \qquad
 P^{\alpha \dot\alpha}\equiv P^\mu \sigma_\mu^{\alpha\dot\alpha}
=\PBK{1 \\ z}^\alpha \PBK{1 \\  \bar z}^{\dot\alpha} e^{  \p_\Delta} ~.
 \ee
where $\alpha,\dot\alpha=1,2$. Applying \eqref{Qtsf} to superfields \eqref{superfd1} and \eqref{superfd2}, we obtain the susy transformation rules of all the component  operators, as shown in  \eqref{Qsusy} and \eqref{Qsusy2}.

\section{Useful Integrals}\label{integral}
In this appendix we derive  some useful integrals which are crucial for shadow transformation. 
\subsection{Seed formula}
We first want to compute \footnote{ Note that $z,\bar z$ are complex conjugate of each other in this appendix.}
\be
I(z_1,z_2) =\int d^2z \frac{1}{(z-z_1)(\bz-\bz_2)}
=\int d^2z \frac{(\bz-\bz_1)( z- z_2)}{|z-z_1|^2|\bz-\bz_2|^2}~.
\ee
Using Feynman parametrization, \footnote{~Namely:
\be\nonumber
\frac{1}{AB}=\int_0^1 du \frac{1}{(uA+(1-u)B)^2}~.
\ee}
we can rewrite the integral as
\beqn
I(z_1,z_2) &=&\int_0^1 du \int d^2z \frac{(\bz-\bz_1)( z- z_2)}{  (u|z-z_1|^2+(1-u)|\bz-\bz_2|^2 )^2}
 \\&=&
\int_0^1 du \int d^2w \frac{ (\bar w-(1-u) \bz_{12})(w+uz_{12})}{(|w|^2+u(1-u) |z_{12}|^2)^2} 
\\&=&
\int_0^1 du \int d^2w \frac{   | w|^2- u(1-u)| \bz_{12} |^2+u\bar w z_{12}-(1-u)  w \bz_{12}}{(|w|^2+u(1-u) |z_{12}|^2)^2} ~.
\eeqn
where we change variable to  $w=z+uz_1+(1-u)z_2$.  In the last expression, the last two terms in the numerator is supposed to have no contribution because they are odd function of $w,\bar w$. The resulting integral is divergent, so we need to regularize it. We analytically continue the dimension from 2 to $d=2+\epsilon$, then the integral becomes
\beqn
I(z_1,z_2) &=&\int_0^1 du \int d\rho \rho^{d-1} V_{d-1} \frac{  \rho^2- u(1-u)| \bz_{12} |^2 }{( \rho ^2+u(1-u) |z_{12}|^2)^2}
  \\ &=&
\frac{2\pi^{d/2}}{\Gamma(d/2)}\int_0^1 du \frac{(d-1)\pi \big(u(1-u)| \bz_{12} |^2\big)^{\frac d2-1} }{2\sin \frac{d\pi}{2}}  \qquad \qquad( \text{for }d<2 )
  \\ &=&
\frac{2^{2-d} \pi ^{\frac{d+3}{2}}  |z_{12}|^{d-1}  }{\Gamma  (\frac{d-1}{2} )\sin \frac{d\pi}{2}}
  \\ &\xrightarrow{\epsilon\to 0}&
 -\pi \Big( \frac{2}{\epsilon}+\ln \pi+\gamma_E+\ln |z_{12}|^2+\mathcal O(\epsilon)\Big) ~,
 \label{seedintegral}
\eeqn
 where $V_{d-1}$ is the area of of $d-1$-dimensional sphere $S^{d-1}$.  We can regard it as a seed formula and generate many other integrals by taking derivative with respect to $z_1,z_2$. 

\subsection{General formula}
The integrals relevant for shadow transformation generally take the  following form
\be\label{IsAB}
I_{s }^{(A,B)}(w,z')=\int d^2 z\; (z-w)^A (\bar z-\bar w)^B \frac{(\bar z-\bar z')^s}{z-z'}~.
\ee
Taking derivative with respect to $\bar z'$, we get 
\footnote{~Note that $\p_{\bar z'}$ can hit $1/(z-z') $ and generates delta-function $\delta^{(2)}(z-z')$, but this has vanishing contribution because of $(\bar z-\bar z')^s$ as long as $s>0$. }
\be
\p_{\bar z'} I_{s }^{(A,B)}(w,z')=-s I_{s-1 }^{(A,B)}(w,z'),\qquad s>0~.
\ee
Using this formula iteratively, we get
\be\label{Isrecur}
\p_{\bar z'}^s I_{s }^{(A,B)}(w,z')= (-1)^s s! I_{0 }^{(A,B)}(w,z'),\qquad s\in \mathbb N~.
\ee
Furthermore, we have 
\beqn
\p_{\bar z'}I_{0}^{(A,B)}(w,z')
&=&
\int d^2 z\; (z-w)^A (\bar z-\bar w)^B\p_{\bar z'} \frac{1}{z-z'}
\\&=&
\int d^2 z\; (z-w)^A (\bar z-\bar w)^B (-\pi) \delta^{(2)}(z-z')
\\&=&
  (-\pi)  (z'-w)^A (\bar z'-\bar w)^B ~,
\label{I0res}
 \eeqn
 where we used the formulae
 \be
 \p_{\bar z}\frac{1}{z}= \pi \delta^{(2)}(z), \quad 
 \delta^{(2)}(z)=\delta(x) \delta(y), \quad d^2z=dx dy, \quad z=x+{\rm i}y, \quad
 \int d^2  z \;   f(z)  \delta^{(2)}(z)=f(0)~.\quad
 \ee

Making use of translational invariance and $SL(2,\mathbb C)$ invariance, the integral \eqref{IsAB} is supposed to have the following structure \footnote{~This might fail if the exponent is zero. More specifically, if $A=0$, we may have $\log (z'-w)$ instead of $(z'-w)^0$. And similarily for anti-holomorphic part $\bar z'-\bar w$. Besides, there may be also some contact-terms in the form $\p^\#\delta(z'-w)$, but this can only contribute in the coincident limit.  They have no contribution in our discussion of Ward-identities as long as all the points are distinct, so we ignore this type of contribution. }
\be
I_{s }^{(A,B)}(w,z')=c_{s }^{(A,B)} \times  (z'-w)^A (\bar z'-\bar w)^{B+s+1}~,
\ee
where $c_{s }^{(A,B)}$ is a constant. We can use   relations \eqref{Isrecur} and \eqref{I0res} to determine the constant.  As a consequence, we find 
\be\label{IsABres}
I_{s }^{(A,B)}(w,z')=\int d^2 z\; (z-w)^A (\bar z-\bar w)^B \frac{(\bar z-\bar z')^s}{z-z'}
=\frac{(-1)^s(-\pi) s!}{(B+1) (B+2) \cdots (B+s+1)} (z'-w)^A (\bar z'-\bar w)^{B+s+1}~.
\ee

\subsection{Some examples}

In this subsection we would like to evaluate some integrals explicitly which are relevant for the discussion in the main body. We will use the techniques established in the previous two subsections which are independent. The agreement of different approaches justify our prescription in computing the integral. 

We first would like to compute the following integral:
\beqn
I_1^{(0,-4)}(w,z')&=&   \int d^2 z    \frac{1}{(\bz-\bar w)^4}\frac{ \bar z- \bar z'   }{z-z'}  
=    \int d^2 z   \frac{1}{(\bz-\bar w)^3(z-z')}  
+ (\bar w-\bz') \int d^2 z   \frac{1}{(\bz-\bar w)^4(z-z')}  
\nonumber\\ &=&
\Big(  \frac12\p_{\bar w} ^2 +\frac16 (\bar w-\bz')  \p_{\bar w} ^3 \Big) 
  \int d^2 z   \frac{1}{(\bz-\bar w) (z-z')}  
\nonumber \\ &=&
\frac{\pi}{6} \frac{1}{(\bar w -\bz')^2}~,
\label{I104}
\eeqn
where in the last equality we used \eqref{seedintegral}. This agrees with the general formula \eqref{IsABres}. 
 
 The second integral of our interest is
 \beqn
I_2^{(0,-4)}(w,z')&=&   \int d^2 z    \frac{1}{(\bz-\bar w)^4}\frac{ (\bar z- \bar z'  )^2 }{z-z'}  
= \frac16\p_{\bar w} ^3  \int d^2 z    \frac{1}{(\bz-\bar w) }\frac{ (\bar z- \bar z'  )^2 }{z-z'}  
 \\&=& 
  \frac16\p_{\bar w} ^3    \int d^2 z    \frac{(\bz-\bar\eta)^2}{(z-  \eta)   \bz}  , \qquad\qquad \eta = z' -w
  \\&=& 
  \frac16\p_{\bar w} ^3    \int d^2 z   
  \Big[  \frac{\bz}{z-\eta}-\frac{2\eta}{z-\eta}+\bar\eta^2 \frac{1}{\bz(z-\eta)} \Big]
    \\&=& 
 - \frac16\p_{\bar \eta} ^3    \int d^2 z   
  \Big[  \frac{\bz+\bar\eta}{z }-\frac{2\eta}{z }+\bar\eta^2 \frac{1}{\bz(z-\eta)} \Big]
      \\&=& 
       \frac{\pi}{3} \frac{1}{\bz'-\bar w}~,
       \label{I204}
\eeqn
 where  we used the translational invariance several times and  \eqref{seedintegral} in the last equality.  
 This is again consistent with  the general formula \eqref{IsABres}.  One can also verify  the relation $\p_{\bar z'}I_2^{(0,-4)}(w,z')=-2I_1^{(0,-4)}(w,z')$ as expected.  These two examples $I_1^{(0,-4)},I_2^{(0,-4)}$ are directly related to the derivation of stress tensor Ward identity in \eqref{stresstensorWd}.
  
Therefore we have checked our general formula \eqref{IsABres} in various examples.

\section{General celestial OPEs in EFT} \label{opeEFT}

We will consider the   collinear limit  in effective field theory. In particular, we consider the collinear limit arising from the cubic   vertex  involving massless particles with helicity $s_1,s_2,s_3$.  In this limit, the amplitude satisfies
\be
A_n(1^{s_1},2^{s_2}, \cdots) \xrightarrow{p_1\//\!\// p_2}   \sum_{s_3} \Split(1^{s_1}+2^{s_2}\to P^{-s_3} ) 
 A_{n-1}(P^{-s_3}, \cdots)~,
\ee
where $P=p_1+p_2$ and  the dots represent the rest of  $n-2$ particles which are not relevant here. In the collinear limit, the scattering amplitude is  also supposed to factorize as follows:
\be
A_n(1^{s_1},2^{s_2}, \cdots) \xrightarrow{p_1\//\!\// p_2}   \sum_{s_3} A_3(1^{s_1}+2^{s_2}, -P^{ s_3} ) \frac{1}{P^2}
 A_{n-1}(P^{-s_3}, \cdots)~.
\ee
Therefore we have 
\beqn
 \Split(1^{s_1}+2^{s_2}\to P^{-s_3} )  
 &=&
 A_3(1^{s_1},2^{s_2}, -P^{s_3}) \frac{1}{P^2}
 \propto   \frac{1}{\EV{12}[12]} [12]^{s-2s_3}[1P]^{s-2s_2}[P2]^{s-2s_1}~, \qquad
 \eeqn
where we use the property that the three-point amplitude $ A_3$ is uniquely fixed by the symmetry and locality \cite{Elvang:2013cua}. Also we assume that $s\equiv s_1+s_2+s_3 \ge 0$, so locality guarantees that only square brackets appear in the three-point amplitude. And $P^2=(p_1+p_2)^2=2 p_1\cdot p_2
\propto \EV{12}[12]$.  Note that we will not keep track of any overall constant as they can  be absorbed into the effective coupling which we also don't write down explicitly. 

Because of momentum conversation $P=p_1+p_2$, we actually have (for $s\ge 0$)
\footnote{~To be more rigorous, one needs to deviate from the strict collinear limit a little bit. Since we are only interested in the leading singular term, this infinitesimal deviation is not important.} 
\beqn
\EV{ij}=0\qquad &&\rightarrow\qquad \lambda_P =\alpha_1\lambda_1=\alpha_2\lambda_2~, \\
P=p_1+p_2\qquad &&\rightarrow\qquad \lambda_P \tilde \lambda_P =\lambda_1 \tilde \lambda_1+\lambda_2 \tilde \lambda_2  ~.
\eeqn
This gives
\be
\tilde\lambda_P=\frac{1}{\alpha_1}\tilde\lambda_1+\frac{1}{\alpha_2}\tilde\lambda_2~,
\ee
 and thus 
 \be
[1P]=\frac{1}{\alpha_2}[12]~,\qquad  [P2]=\frac{1}{\alpha_1}[12]~,
 \ee

Furthermore for physical momentum $\tilde \lambda=\lambda^*$, we thus have $1/\alpha_1^2+1/\alpha^2_2=1$ which enables us to  set $1/\alpha_1 =\sqrt{x}, 1/\alpha_2 =\sqrt{1-x}$. As a consequence,  $p_1=xP, \; p_2=(1-x)P$ and  $\omega_1=x \omega_P, \; \omega_2=(1-x)\omega_P$.\footnote{~Here $\omega$ is the energy along the  null direction such that $p_i^\mu=\eta_i\omega_i q_i^\mu$ and $q=\Big(1+z \bar z ,   z+\bar z ,  -i(z-\bar z),   1-z \bar z\Big)$. In the collinear limit, $q_i$'s are approximately the same.}

With these relation, the split factor can be evaluated explicitly 
\beqn
 \Split(1^{s_1}+2^{s_2}\to P^{-s_3} )  
 &\propto&
 \frac{[12]^{s-1}}{\EV{12} }   \Bigg( \frac{[1P]}{[12]}  \Bigg)^{s-2s_2}   \Bigg( \frac{[P2]}{[12]} \Bigg )^{s-2s_1}
  \\ &=&
 \frac{[12]^{s-1}}{\EV{12} }   ( \sqrt x  )^{s-2s_2}   ( \sqrt{1-x} )^{s-2s_1}
   \\ &\propto&
 \frac{ \bz_{12}^{s-1}}{z_{12} }  \omega_1^{s_2+s_3-1}  \omega_2^{s_1+s_3-1} \omega_P^{-s_3}~,
 \label{split}
  \eeqn
 where we used $\EV{ij}=
-2 \sqrt{\omega_i \omega_j} \; z_{ij}, \;
{}[ij] =
2  \eta_i \eta_j \sqrt{\omega_i \omega_j} \; \bar z_{ij}$, 
 $
 z_{ij}=z_i-z_j,
\bar z_{ij}=\bar z_i- \bar z_j$ and $\eta_i=\pm 1$ distinguishes outgoing/incoming particles. 
 
 We are however interested in the celestial amplitude which is defined as 
  \be\label{MellinTsf}
 \mathcal M_n(\Delta_i,J_i,  z_i, \bar z_i)
 =\Big(\prod_{j=1}^n  \int_0^\infty d\omega_j\; \omega_j^{\Delta_j -1}  \Big)  \mathcal A_n( J_i, p^\mu_i)~,
 \ee
 where $J_i$    is the spin of the operator in 2d, while in 4d it is the helicity of the particle, namely $J_i=s_i$.
 The   celestial amplitude can be regarded as a conformal correlator on the celestial sphere
\be
 \mathcal M_n(\Delta_i,J_i,  z_i, \bar z_i)=\EV{\cO^{\eta_1}_{\Delta_1,J_1}(z_1,\bar z_1) \cdots  \cO^{\eta_n}_{\Delta_n,J_n}(z_n,\bar z_n) }~.
\ee 

To finally obtain the OPE, we just need to perform   the Mellin transformation for \eqref{split}.
In particular, we have    the Mellin transformation of  $\omega_1^\alpha \omega_2^\beta (\omega_1+\omega_2)^\gamma$ in the split factor:
\beqn\label{collinearBfcn}
&& \int_0^\infty \!\!   d\omega_2\;  \omega_2^{\Delta_2-1} \int_0^\infty  \!\!    d\omega_1\; \omega_1^{\Delta_1-1} 
\omega_1^\alpha \omega_2^\beta (\omega_1+\omega_2)^\gamma \; f(\omega_1+\omega_2)
= B\Big(\Delta_1+\alpha,\Delta_2+\beta \Big)    \!\!   
  \int_0^\infty d\omega \; \omega^{ \Delta_P  -1}  \; f(\omega)~,
  \qquad\nonumber\\
\eeqn
where we used \eqref{betafcn} and  $\Delta_P=\Delta_1+\Delta_2+\alpha+\beta+\gamma$.

With this formula, we finally obtain  tree level leading order  celestial OPE arsing from cubic vertex in EFT:
\be
\cO_{\Delta_1, J_1}(z_1, \bz_1)\cO_{\Delta_2, J_2}(z_2, \bz_2)
\sim  \frac{ \bz_{12}^{J_1+J_2+J_3-1}}{z_{12} }
B\Big(\Delta_1+J_2+J_3-1,\Delta_2+J_1+J_3-1 \Big) 
\cO_{\Delta_3, -J_3}(z_2, \bz_2)~, \qquad
\ee
where $\Delta_3=\Delta_1+\Delta_2+J_1+J_2+J_3-2$.

\section{Soft photon theorems with magnetic corrections    } \label{softphoton}

 The leading and sub-leading soft photon theorems state  that  
  \be
 \lim_{p \to 0}M_{n+1}(p^s, p_1, p_2, \cdots, p_n)=\Big( S^{(0)}_s  + S^{(1)}_s  +\cdots \Big) M_n(p_1, p_2, \cdots p_n)~,
 \ee
 with soft factors
 \be
S^{(0)}_s =\sum_{k=1}^n \eta_k \frac{ (e_k    \epsilon_s+g_k \tilde \epsilon_s )\cdot p_k   }{p\cdot p_k}~,
  \qquad
S^{(1)}_s =\sum_{k=1}^n  i \eta_k \frac{ (e_k    \epsilon_s ^\mu+g_k \tilde \epsilon_s^\mu )  p ^\nu J_{k} {}_{\mu\nu}  }{p\cdot p_k}~,
 \ee
 where $\eta_k=\pm 1$ for out-going/in-coming particle, $s$ is the helicity of the soft photon, $e_k,g_k$ are the electric and magnetic charges   of $k$-th  matter particle, while $q^\mu_k, J_k{}_{\mu\nu}$ represent    momentum and angular momentum. In the absence of magnetic charge, these are the Weinberg's soft photon theorem and Low's sub-leading soft photon theorem. The role of magnetic charges can be incorporated through the electro-magnetic duality transformation, which can be regarded as a phase rotation of couplings. The invariance of soft factors under the gauge transformation   $ \epsilon_s^\mu\to  \epsilon_s^\mu+p^\mu,\; \tilde  \epsilon_s^\mu\to  \tilde \epsilon_s^\mu+p^\mu$  is equivalent to the conservation of electric and magnetic charges. 
 
 It is convenient to parametrize the momentum in terms of celestial coordinates as follows:
  \be
p^\mu  =\omega q^\mu  , \qquad q^\mu  =\Big(1+z \bar z ,\quad  z+\bar z ,\quad -i(z-\bar z),\quad  1-z \bar z\Big)~.
 \ee
Then the (electric) polarization vectors can be chosen as:
\be
 \epsilon_+^\mu (p)=\frac{1}{\sqrt 2 } \p_z q^\mu =(\bar z,\, 1, \, -i,\, -\bar z)~, \qquad
 \epsilon_-^\mu (p)=\frac{1}{\sqrt 2 }  \p_{\bar z} q^\mu =(  z,\, 1, \, i,\, -  z)~,
\ee
satisfying 
\be\label{pole}
 \epsilon_+\cdot p= \epsilon_{ -}\cdot p=0~,\qquad  \epsilon_+\cdot \epsilon_+= \epsilon_{-}\cdot \epsilon_{-}=0~, \qquad 
\epsilon_+\cdot    \epsilon_{-}=1~.
 \ee
We also need  the magnetic polarization vectors which are defined as follows~\cite{Terning:2018udc} 
\be
\tilde \epsilon_\pm{}_\mu (p)=\frac{\varepsilon_{\mu\nu\rho\sigma} n^\nu p^\rho   \epsilon_\pm^\sigma (p)  }{p \cdot n }~,
\ee 
where $n^\rho$ is an arbitrary reference vector. Explicit computations show that
\be
\tilde \epsilon_+^\mu (p)= i  \epsilon_+^\mu (p) +r \, p^\mu~, \qquad
\tilde \epsilon_-^\mu (p)=- i  \epsilon_ -^\mu (p) +\bar r\, p^\mu~, \qquad
\ee
where $r$ is function of $n^\mu$ and $p^\mu$.  Since  $rp^\mu, \bar r p^\mu$ just correspond to gauge transformations, they can be dropped out in the polarizations. As a result, the on-shell amplitudes do not depend on the  reference vector  $n^\mu$. The magnetic polarizations satisfy the same relation as the electric ones in \eqref{pole}.
 
Using   celestial coordinates,   the soft factors  for  positive soft photon are given by 
\be\label{softphotontheorem}
 \qquad S^{(0)}_+=\sum_{k=1}^n \eta_k \frac{Q_k}{\omega(z-z_k)}~, \quad 
S^{(1)}_+=\sum_{k=1}^n \eta_k \frac{Q_k}{  z-z_k }
\Big(\frac{J_k}{\omega_k} +\p_{\omega_k}+\frac{\bz-\bz_k}{\omega_k} \bar\p_k\Big) ~, \quad 
\qquad
Q_k =e_k+i g_k~.
\ee
For negative soft photon, the soft factors  are given by  its complex conjugate. 

 \bibliographystyle{JHEP} 
 
\bibliography{BMS.bib} 
  
\end{document}